\documentclass{article}

\usepackage[preprint]{neurips_2023}
\usepackage[utf8]{inputenc} 
\usepackage[T1]{fontenc}    
\usepackage{hyperref}       
\usepackage{url}            
\usepackage{booktabs}       
\usepackage{amsfonts}       
\usepackage{nicefrac}
\usepackage{microtype}      
\usepackage{xcolor}         

\usepackage{amsthm}
\usepackage{amsmath}
\usepackage{graphicx}
\usepackage{subcaption}
\usepackage{hyperref}
\usepackage{tabularx}
\usepackage[parfill]{parskip}
\usepackage{cleveref}
\usepackage{setspace}      
\usepackage{threeparttable}
\usepackage{caption}
\usepackage{amssymb}

\title{Neuroimaging Meta Regression for Coordinate Based Meta Analysis Data with a Spatial Model}

\author{\href{https://orcid.org/0000-0002-9741-0051}{\includegraphics[scale=0.06]{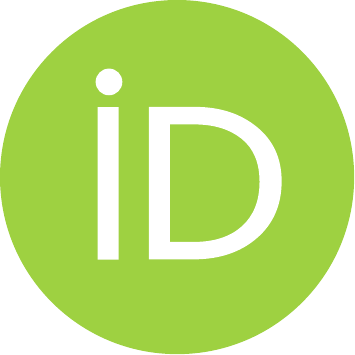}\hspace{1mm}Yifan Yu}\\
	Oxford Big Data Institute\\
	University of Oxford\\
	Oxford, UK \\
	\And
	\href{https://orcid.org/0000-0002-7679-1385}{\includegraphics[scale=0.06]{figures/orcid.pdf}\hspace{1mm}Rosario Pintos Lobo} \\
	Department of Psychology\\
	Florida International University\\
	Miami, FL, USA \\
        \And
	Michael Cody Riedel \\
	Department of Physics\\
	Florida International University\\
	Miami, FL, USA \\
        \And
        \href{https://orcid.org/0000-0002-7796-8795}{\includegraphics[scale=0.06]{figures/orcid.pdf}\hspace{1mm}Katherine Bottenhorn} \\
        Department of Population and Public Health Sciences \\
        University of Southern California \\
        Los Angeles, CA, USA \\
	\And
	\href{https://orcid.org/0000-0003-3379-8744}{\includegraphics[scale=0.06]{figures/orcid.pdf}\hspace{1mm}Angela R. Laird} \\
	Center for Imaging Science\\
	Florida International University\\
	Miami, FL, USA \\
	\And
	\href{https://orcid.org/0000-0002-4516-5103}{\includegraphics[scale=0.06]{figures/orcid.pdf}\hspace{1mm}Thomas E.~Nichols}\thanks{To whom correspondence should be addressed.} \\
	Oxford Big Data Institute\\
	University of Oxford\\
	Oxford, UK \\
	\texttt{thomas.nichols@bdi.ox.ac.uk} \\
}

\begin{document}

\maketitle

\begin{abstract}
  Coordinate-based meta-analysis combines evidence from a collection of Neuroimaging studies to estimate brain activation. In such analyses, a key practical challenge is to find a computationally efficient approach with good statistical interpretability to model the locations of activation foci. In this article, we propose a generative coordinate-based meta-regression (CBMR) framework to approximate smooth activation intensity function and investigate the effect of study-level covariates (e.g., year of publication, sample size). We employ spline parameterization to model spatial structure of brain activation and consider four stochastic models for modelling the random variation in foci. To examine the validity of CBMR, we estimate brain activation on $20$ meta-analytic datasets, conduct spatial homogeneity tests at voxel level, and compare to results generated by existing kernel-based approaches. 
\end{abstract}

\section{Introduction}
Functional neuroimaging includes a number of techniques to image brain activity, including Positron emission tomography (PET) and functional magnetic resonance imaging (fMRI). Starting three decades ago, PET studies were used to compare brain activity between rest and experimental conditions, producing maps of ``activation", images of statistics measuring the strength of the experimental effect. Especially in the last two decades, the literature of fMRI activations has grown rapidly, which motivates a need to integrate findings, establish consistency and heterogeneity across independent but related studies. However in both PET and fMRI studies, the validity is challenged by common drawbacks of small sample size, high prevalence of false positives (approximately $10-20\%$ of reported foci in publications are false positives (\citealp{wager2007meta}), as well as significant heterogeneity among studies and unreliable inference due to their diversity in measurements and types of analysis (\citealp{samartsidis2017coordinate}). Meta-analysis is an essential tool to address these limitations and improve statistical power by pooling evidence from multiple studies and providing insight into consistent results. 

There are also applications of neuroimaging meta-analysis to resting-state fMRI and structural analysis using voxel-based morphometry. Going forward we will only reference fMRI, but note that the application extends to other types of data. Meta-analysis is classified into two categories in neuroimaging research: image-based meta-analysis (IBMA) which uses the 3D statistic maps of original studies and coordinate-based meta-analysis (CBMA) which uses the reported spatial coordinates of activation foci in a standard MNI or Talairach space. Ideally, only IBMA would be used, as there is substantial information loss by only using activation foci as compared to full statistics maps, and further accuracy loss occurs when deactivation foci are ignored (\citealp{salimi2009meta}). However, while it is now more common to share entire statistical maps in published studies, historically this has not been the case (\citealp{salimi2009meta}) and there exists large-scale coordinate databases (e.g., BrainMap (\citealp{laird2005brainmap}), Neurosynth (\citealp{yarkoni2011large})). Hence, CBMA is still the predominant approach for neuroimaging meta-analysis. 

To identify consensus in brain regions with consistent activation across studies, researchers have developed a variety of CBMA methods, which are either kernel-based or model-based. Among those kernel-based methods, activation likelihood estimation (ALE, with a Gaussian kernel), multilevel kernel density analysis (MKDA, with a uniform sphere) and signed differential mapping (SDM, with a Gaussian kernel scaled by effect size) are commonly used (\citealp{turkeltaub2002meta,eickhoff2012activation,wager2007meta,radua2012new}). None of the three methods is based on a formal statistical model, however, all are able to obtain statistical inferences by reference to a null hypothesis of total random arrangement of the foci (\citealp{samartsidis2017coordinate}). Voxels with significant p-values are considered to be the regions of consistent activation. Multiple testing corrected inferences are made by controlling the family wise error rate using the null maximum distribution (\citealp{westfall1993resampling}) or the false discovery rate (FDR) (Benjamini-Hochberg procedure). However, kernel-based methods lack interpretability as mass-univariate approaches rather than explicit probabilistic models, they do not generally allow group comparison, do not model the spatial dependence of activation foci, as well as cannot accommodate study-level covariates to conduct a meta-regression (\citealp{samartsidis2019bayesian}). 

Bayesian model-based methods address these limitations, and are categorised into parametric spatial point process models (\citealp{kang2011meta, montagna2018spatial, samartsidis2019bayesian}) and non-parametric Bayesian models (\citealp{yue2012meta, kang2014bayesian}). They use explicit generative models for the data with testable assumptions. Although they generally provide advances in interpretability and accuracy over kernel-based methods, they are computationally intensive approaches and generally require parallel computing on GPUs (\citealp{samartsidis2019bayesian}), and only some approaches can conduct meta-regression to estimate the effect of study-level covariates. Further, it can be more challenging for practitioners to interpret the spatial posterior intensity functions and utilise spatial Bayesian models in practice. 

In this work, we investigate classical frequentist models that explicitly account for the spatial structure of the distribution of activation foci. Specifically, we focus on developing a spatial model that takes the form of a generalised linear model, where we make use of a spline parameterization to induce a smooth response and model the entire image jointly, allow for image-wise study-level regressors and consider different stochastic models to find the most accurate but parsimonious fit. Although Poisson is the classic distribution describing independent foci counts, we have previously found evidence of over-dispersion (\citealp{samartsidis2017estimating}), and thus we further explore a Negative Binomial model, a Clustered Negative Binomial model and a Quasi-Poisson model to allow excess variation in counts data.  

Our work draws on the existing methods for CBMA, while introducing key innovations. From the Bayesian work, we take the idea of explicit spatial models; from the kernel methods, we take the idea of fixing the degree of spatial smoothness. The contribution of this meta-regression model is both methodological and practical, it provides a generative regression model that estimates a smooth intensity function and can have study-level regressors. Meanwhile, using a crucial memory-saving model factorisation, it is also a computationally efficient alternative to the existing Bayesian spatial regression models and provides an accurate estimation of the intensity function. While our method is suitable for any CBMA data, we are particularly motivated by studies of cognition. Cognition encompasses various mental processes, including perception, intelligence, problem solving, social interactions, and can be affected by substance use. We demonstrate this meta-regression framework on previously published meta-analyses of $20$ cognitive and psychological tasks, allowing generalised linear hypothesis testing on spatial effect, as well as inference on the effect of study-level covariates. 

In the reminder of this work, we present the proposed meta-regression framework, discuss model factorisation and optimisation procedures, as well as inferences on meta-regression outcomes via statistical tests in \Cref{section: methods}. Then we explain experiment settings in \Cref{section: Experiments} and explore different variants of stochastic models on the $20$ meta-analytic datasets and describe multiple goodness-of-fit statistics to identify the most accurate model, establish valid FPR control via Monte Carlo simulation under the null hypothesis of spatial homogeneity, followed by a comparison of homogeneity test with kernel methods in \Cref{section: Results}. Finally, \Cref{section: Discussion} summarises our findings and potential extension of this meta-regression framework in the future. 

\section{Methods}
\label{section: methods}

Generalised linear models are described in terms of their stochastic and deterministic components. Our 
 deterministic model has a regression structure with a spatial component utilising a spline parameterization and study-level covariate component. For the stochastic model, we consider multiple models motivated by CBMA data characteristics. We then propose a model factorisation approach to make our methods scalable, before outlining a general inference framework.

\subsection{Deterministic model}

\subsubsection{Generic regression structure}
\label{subsubsection: Generic regression structure} 
Assume there are $N$ voxels in each of $M$ studies, and then our CBMA data at voxel $j$ for study $i$ is the voxelwise count of foci $Y_{ij}$, collected for study $i$ as the N-vector $Y_i = \left[Y_{i1}, Y_{i2}, \cdots, Y_{iN} \right]^\top$. We generate a spatial design matrix $X (N \times P)$ with $P$ cubic B-spline bases (more details to follow in \Cref{subsubsection: Spline parameterization}) and construct study-level covariates matrix $Z (M \times R)$ by extracting $R$ study-level covariates from each of $M$ studies. For the CBMA framework, the central object of interest is the voxelwise intensity function for study $i$, which considers both effects of smooth spatial bases and study-level covariates. In this setting, it is most concise to write the model for study $i$ as 
\begin{equation}
\label{equation: deterministic equation}
    \log(\mu_i) = \log\left[\mathbb{E}(Y_i)\right] = X \beta + (Z_i \gamma) \mathbf{1}_N
\end{equation}
where $\beta (P \times 1)$ and $\gamma (R \times 1)$ are regression coefficients for spatial bases $X$ and study-level covariates $Z$ respectively, $Z_i$ is the $i^{th}$ row of study-level regressors $Z$, $\mathbf{1}_N$ is a $N$-vector of $1$'s; and the estimated intensity is captured via $\mu_{ij}$ for studies $i=1,...,M$ and voxels $j=1,...,N$, collected for study $i$ as the N-vector $\mu_i = \left[\mu_{i1}, \mu_{i2}, \cdots \mu_{iN} \right]^\top$. This model is identifiable as long as we ensure each covariate variable is mean zero, letting $X$ capture the overall mean. The GLM for all voxels in all $M$ studies is then
\begin{equation}
\label{eqn: full model}
    \log \left[\mathbb{E}(Y)\right] = (\mathbf{1}_M \otimes X) \beta + (Z \otimes \mathbf{1}_N) \gamma
\end{equation}
where $Y=[Y_1, Y_2, \cdots, Y_M]^\top$ is a $(M \times N)-$vector, containing voxelwise foci count for all of $M$ studies, and $\otimes$ is the Kronecker product. This formulation has millions of rows ($MN$) and the spatial design matrix has billions of entries ($MN \times P$). In consideration of implementation complexity and memory requirement, we will propose a simplified reformulation of this GLM in \Cref{subsection: Model factorisation}. 

\subsubsection{Spline parameterization}
\label{subsubsection: Spline parameterization}
\begin{figure}[h]
\includegraphics[width=11cm]{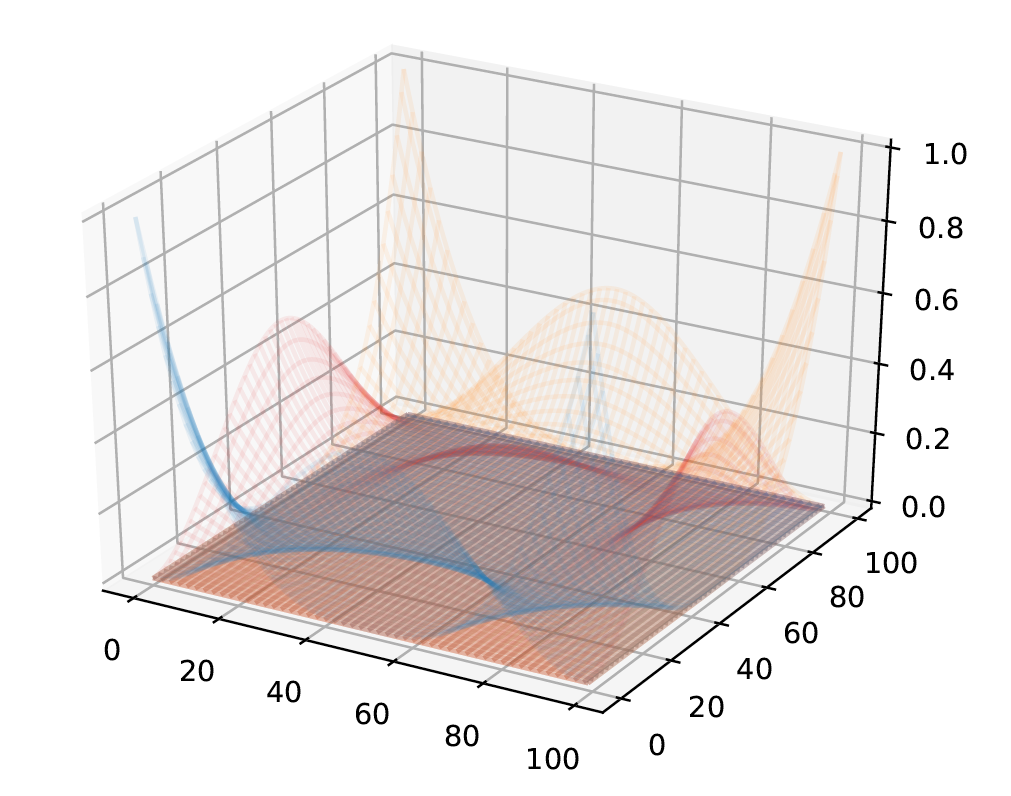}
\centering
\caption{Illustration of tensor product of $2D$ spline bases (with equal knots spacing)}
\label{figure: tensor product of 2D spline bases}
\end{figure}

Previous work on spatial point process modelling of CBMA data have treated each study's foci as realisation of a doubly-stochastic Poisson process, also known as a Cox process. In some of that work, the log intensity function is parameterised by superimposed Gaussian kernel basis functions (\citealp{montagna2018spatial}), while in others, the log intensity is a Gaussian process (\citealp{samartsidis2019bayesian}). Here, we propose tensor product of cubic B-spline basis for modelling the spatial intensity, as its smoothness, stability and local support make it an ideal spatial basis for CBMA application. A 1-dimensional cubic B-spline is a piecewise polynomial of order $k=3$, where pre-specified knots $T=(t_0, t_1, \cdots, t_n)$ determines the parameterization of basis functions as the intersections of polynomial sections. The order $k$ B-spline basis functions $B_{ik}$ are defined by recurrence relations, for $i=0, 1, \cdots, n$,
\begin{equation}
\begin{split}
    B_{i1} &= 
    \begin{cases}
      1 \; \text{for } t_i \leq t < t_{i+1}\\
      0 \; \text{otherwise}
    \end{cases} \; \text{for} \; k=1 \, , \\
    B_{i,k}(t) &= \frac{t-t_i}{t_{i+k-1}-t_i} B_{i, k-1}(t) + \frac{t_{i+k}-t}{t_{i+k}-t_{i+1}} B_{i+1, k-1}(t) \; \text{for} \; k>1 \, .
\end{split}
\end{equation}
The B-spline curve is a linear combination of the B-spline basis function $B_{ik}$. For our $3D$ lattice, assume there are $v_x$ voxels along x direction, the coefficients of $v_x$ voxels evaluated at each of $n_x$ B-spline bases construct a coefficient matrix $C_x$ (size $v_x \times n_x$). Similarly, there exist another two coefficient matrices $C_y$ and $C_z$ (size $v_y \times n_y$ and $v_z \times n_z$) along y and z direction. The whole coefficient matrix $C$ of 3-dimensional B-spline bases is constructed by taking tensor product of the $3$ coefficient matrices (see \Cref{figure: tensor product of 2D spline bases} for a 2D illustration),
\begin{equation}
    C = C_x \otimes C_y \otimes C_z
\end{equation}
The matrix of $C$ is $(v_x v_y v_z) \times (n_x n_y n_z)$, and is the basis for the entire $3D$ volume, while the analysis is based on a brain mask of $N$ voxels. The design matrix $X$ is obtained from $C$ after a three-step process: First, rows corresponding to voxels outside the brain mask are removed; then, columns are removed if they correspond to weakly supported B-spline bases (a B-spline basis is regarded as ``weakly supported" if its maximum value of coefficients evaluated at each voxel is below $0.1$). Finally, the rows are re-normalised (sum to $1$) to preserve the property of ``partition of unity" of spline bases. 

We define our cubic B-spline bases with equally spaced knots on $x,y,z$ dimension, and thus we parameterise the level of spatial smoothness by the knot spacing. Larger knots spacing, smaller basis, and greater smoothness; conversely, closer knots, larger basis, and greater ability to represent fine details. Conceptually, more flexible parameterizations would allow arbitrary knots locations, but with the consideration of minimising computational complexity, we fix the design matrix $X$ based on pre-specified knots spacing according to prior knowledge. While other spline applications use a dense array of knots and then control smoothness with a roughness penalty, the computational and memory requirements of our spatial model demand that we judiciously select the coarsest spline spacing consistent with our application.

\subsection{Stochastic model}
\label{subsection: Stochastic model}
Assumed stochastic behaviours of CBMA foci data determine the form of statistical likelihood we use. We consider a set of four stochastic models for the distribution of foci counts at voxel level. All of our models take the form of generalised linear models, where inhomogeneous intensity at each voxel is captured by the spline bases and any study-level covariate (as per \Cref{eqn: full model}). We fit our model either by maximising log-likelihood function iteratively via L-BFGS algorithm or iteratively re-weighted least squares (IRLS) for Quasi-likelihood models. To identify the most accurate but parsimonious model, we will elaborate our meta-regression framework for these models and illustrate their strengths and limitations.

\subsubsection{Poisson model} 
\label{subsubsection: Poisson model}
In practice, the count of foci $Y_{ij}$ (for studies $i=1,\cdots, M$, voxels $j=1,\cdots, N$) is only ever $0$ or $1$, which strictly indicates a Binomial model. However, inspired by previous success with Poisson point process, and accuracy of Poisson approximation for low-rate Binomial data (\citealp{eisenberg1966general}), we consider a Poisson model. 

If foci arise from a realisation of a (continuous) inhomogeneous Poisson process, the (discrete) voxel-wise counts will be independently distributed as Poisson random variables, with rate equal to the integral of the (true, unobserved, continuous) intensity function over each voxel. As the sum of multiple independent Poisson random variables is also Poisson, also gives rise to a practical consequence that it is equivalent to either model the set of $M$ study-level counts or the summed counts at each voxel. Following the deterministic structure outlined in \Cref{equation: deterministic equation}, the intensity for voxel $j$ in study $i$ is
\begin{equation}
    \begin{split}
        \mathbb{E}[Y_{ij}] &= \mu_{ij}  \\
        \log(\mu_{ij}) &= \eta_{ij} = x^\top_j \beta + Z_i \gamma
    \end{split}
\end{equation}
where $Y_{ij} \sim \mathrm{Poisson}(\mu_{ij})$, $x^\top_j$ is the $j^{th}$ row of spatial design matrix $X (N \times P)$, and $\beta$ is regression coefficient of spline bases. The data vector $Y$ has length-$(MN)$, which is impractical to represent explicitly. Under the assumption of independence of counts across studies, the likelihood function is exactly same if we model the voxelwise total foci count over studies instead (more details to follow in \ref{appendix: Poisson model} in supplementary material), which gives rise to the modified Poisson model on summed data at voxel $j$ over all studies, $Y_{\centerdot,j}=\sum \limits_{i=1}^M Y_{ij}$,
\begin{equation}
    \begin{split}
    \mathbb{E}[Y_{\centerdot,j}] &= \mu_{\centerdot,j}  \\
    \mu_{\centerdot,j} &= \sum \limits_{i=1}^M \mu_{ij} = \sum \limits_{i=1}^M \exp \left(x_j^\top \beta + Z_i \gamma\right) = \exp(x_j^\top \beta)\left(\sum \limits_{i=1}^M \exp(Z_i \gamma)\right)
    \end{split} 
\label{eq: basic Poisson likelihood}
\end{equation}
where $\mu_{\centerdot,j}=\sum \limits_{i=1}^M \mu_{ij}$ is the expected sum of intensity at voxel $j$ over studies. Under this formulation, the likelihood to be optimised is,
\begin{equation}
    l(\theta) = l(\beta, \gamma) = \sum \limits_{j=1}^N \left[Y_{\centerdot,j}\log(\mu_{\centerdot,j}) - \mu_{\centerdot,j} - \log(Y_{\centerdot,j}!) \right]
\end{equation}

\subsubsection{Negative Binomial model}
\label{subsubsection: Negative Binomial model}
While Poisson model is widely used in the regression of count data, it is recognised that counts often display over-dispersion (the variance of response variable substantially exceeds the mean). Imposition of Poisson model according to an unrealistic assumption (variance equals mean) may underestimate the standard error, and give rise to biased estimation of the regression coefficients. While \citet{barndorff1969negative} proposed a formal definition of spatial Negative Binomial model, it involves Gaussian processes and complexities we sought to avoid. Hence, here we do not pose a formal point process model, but rather simply assert that the count data at each voxel follows a Negative binomial (NB) distribution independently, thus allowing for anticipated excess variance relative to Poisson (\citealp{lawless1987negative}). 

Our NB model uses a single parameter $\alpha$ shared over all studies and all voxels to index variance in excess of Poisson model. For each study $i$ and voxel $j$, let $\lambda_{ij}$ follows a Gamma distribution with mean $\mu_{ij}$ and variance $\alpha \mu_{ij}^2$; then conditioned on $\lambda_{ij}$, let $Y_{ij}$ be Poisson with mean $\lambda_{ij}$. Then it can be shown that the marginal distribution of $Y_{ij}$ follows a NB distribution with probability mass function, 
\begin{equation}
\label{equation: pdf of NB model}
    \mathbb{P}(Y_{ij} = y_{ij}) = \frac{\Gamma(y_{ij}+\alpha^{-1})}{\Gamma(y_{ij}+1) \Gamma(\alpha^{-1})} (\frac{1}{1 + \alpha \mu_{ij}})^{\alpha^{-1}} (\frac{\alpha \mu_{ij}}{1 + \alpha \mu_{ij}})^{y_{ij}}.
\end{equation}
In terms of the success count and probability parameterization, $\text{NB}(r,p)$, we have $Y_{ij} \sim \text{NB}(\alpha^{-1}, \frac{\mu_{ij}}{\alpha^{-1}+\mu_{ij}})$, with mean $\mathbb{E}(Y_{ij}) = \mu_{ij}$ and variance $\mathbb{V}(Y_{ij}) = \mu_{ij} + \alpha \mu_{ij}^2$. Details on derivation of probability density function of NB model can be found in \ref{appendix: Negative Binomial (Poisson-Gamma) model} of the Supplementary material. When $\alpha > 0$, we have Poisson-excess variance of $\alpha \mu_{ij}^2$; or analogous to the coefficient of variation, the coefficient of excess variation is $\sqrt{\alpha \mu_{ij}^2} / \mu_{ij} = \sqrt{\alpha}$, which can be interpreted roughly as the relative excess standard deviation, relative to a Poisson model. 

Again, the data vector is impractical to represent explicitly, but unlike Poisson, the sum of multiple independent NB random variables doesn't follow a NB distribution. Thus, we propose moment matching approach to approximate the mean (first moment) and variance (second moment) of this convolution of NB distributions, which significantly facilitates the simplification of log-likelihood function. Matching the first two moments, the approximate NB distribution of total count of foci over all studies at voxel $j$ is given by $Y_{\centerdot,j} = \sum \limits_{i=1}^M Y_{ij} \sim \mathrm{NB}(r'_j, p'_j)$, where \[r'_j= \frac{\mu_{\centerdot,j}^2}{\alpha \sum \limits_{i=1}^M \mu_{ij}^2}, \;\; p'_j = \frac{\sum \limits_{i=1}^M \mu_{ij}^2}{\alpha^{-1} \mu_{\centerdot,j} + \sum \limits_{i=1}^M \mu_{ij}^2}\] with corresponding excess variance \[\alpha' = \alpha \frac{\sum \limits_{i=1}^M \mu_{ij}^2}{\mu_{\centerdot,j}^2}, \] which gives rise to the simplified NB log-likelihood function,
\begin{equation}
    l(\theta)\approx l (\beta, \alpha') = \sum \limits_{j=1}^N \left[\log\Gamma(Y_{\centerdot,j} + r'_j) - \log\Gamma(Y_{\centerdot,j}+1) - \log\Gamma(r'_j) + r'_j \log{(1-p'_j) + Y_{\centerdot,j} \log{p'_j}} \right]
\end{equation}
Details on derivations of moment matching approach can be found in \ref{appendix: Moment Matching Approach} in the Supplementary material. 

\subsubsection{Clustered Negative Binomial model}
\label{subsubsection: Clustered Negative Binomial model}
While NB model can be regarded as a kind of ``random effects" Poisson model, as developed above, the latent Gamma random variable introduces independent variation at each voxel. We could instead assert that the random (Gamma-distributed) effects are not independent voxelwise effects, but rather latent characteristics of each study, and represent a shared effect over the entire brain for a given study. This is, in fact, the approach used by a Bayesian CBMA method (\citealp{samartsidis2019bayesian}), and in a non-imaging setting, a Poisson-Gamma model for two-stage cluster sampling (\citealp{geoffroy2001poisson}). Therefore, we now consider a third GLM, where at the first stage, we assume each individual study $i$ is sampled with a global latent value $\lambda_i$ from a Gamma distribution with mean $1$ and variance $\alpha$, which accommodates excess variance by dispersion parameter $\alpha$ ($\lambda_i \sim Gamma(\alpha^{-1},\alpha^{-1})$). At the second stage, conditioned on the global variable $\lambda_i$, $Y_{ij}$ are drawn from a Poisson distribution with mean $\lambda_i \mu_{ij}$ ($Y_{ij} \vert \lambda_i \sim \mathrm{Poisson}(\lambda_i \mu_{ij})$), where $\mu_{ij}$ is the expected intensity parameterised by spatial regression parameter $\beta$ and covariates regression parameter $\gamma$. The marginal distribution of $Y_{ij}$ also follows a NB distribution,
\begin{equation}
    \mathbb{P}(Y_{ij} = y_{ij}) = \frac{\Gamma(y_{ij}+\alpha^{-1})}{\Gamma(y_{ij}+1) \Gamma(\alpha^{-1})} (\frac{\alpha^{-1}}{\mu_{ij} + \alpha^{-1}})^{\alpha^{-1}} (\frac{\mu_{ij}}{\mu_{ij} + \alpha^{-1}})^{y_{ij}}
\end{equation}
where $Y_{ij} \sim NB(\alpha^{-1}, \frac{\mu_{ij}}{\alpha^{-1}+\mu_{ij}})$ with mean $\mathbb{E}(Y_{ij}) = \mu_{ij}$ and variance $\mathbb{V}(Y_{ij}) = \mu_{ij} + \alpha \mu_{ij}^2$. Details on derivation of probability density function of clustered NB model can be found in \ref{appendix: Two-stage hierarchy Poisson-Gamma model} in Supplementary material. This two-stage hierarchical Clustered NB model also introduces a covariance structure between foci within a study, which is determined by the expected intensity of the observations as well as the dispersion parameter $\alpha$ (see \ref{appendix: Covariance structure in Clustered NB model} in Supplementary material). The covariance for studies $i$ and $i'$, and distinct voxel $j$ and $j'$ is,
\begin{equation}
    \begin{cases}
      \mathbb{C}(Y_{ij}, Y_{i', j'}) = \alpha \mu_{ij} \mu_{i j'}, \; \text{if} \; i=i'\\
      \mathbb{C}(Y_{ij}, Y_{i', j'}) = 0,  \; \text{if} \; i \neq i'
    \end{cases}
\end{equation}
The log-likelihood is the sum of terms over independent studies,
\begin{equation}
    \begin{split}
        l(\beta, \alpha, \gamma) &= \sum \limits_{i=1}^M \log[f(Y_{i1}, Y_{i2}, \cdots, Y_{iN})] \\
        &= M\alpha^{-1}\log(\alpha^{-1})-M\log\Gamma(\alpha^{-1}) + \sum \limits_{i=1}^M \log\Gamma(Y_{i,\centerdot}+\alpha^{-1}) \\
        &- \sum \limits_{i=1}^M \sum \limits_{j=1}^N \log{Y_{ij}!} - \sum \limits_{i=1}^M (Y_{i,\centerdot} + \alpha^{-1})\log(\mu_{i,\centerdot}+\alpha^{-1}) + \sum \limits_{i=1}^M \sum \limits_{j=1}^N Y_{ij} \log \mu_{ij}
    \end{split}
\end{equation}
where $Y_{i, \centerdot}=\sum \limits_{j=1}^N Y_{ij}$ is the sum of foci within study $i$. One limitation of this model, though, is that it doesn't admit a factorisation and depends on the length-(MN) data vector (see \ref{appendix: Covariance structure in Clustered NB model} in Supplementary material).

Despite a good motivation to induce intra-study dependence, the Clustered NB model depends on the strong assumption that excess variance is captured by the global dispersion $\lambda_i$. If there is voxelwise independent excess variance, the previous NB model will be preferred; we assess this issue below with real data evaluations. 

\subsubsection{Quasi-Poisson model}
\label{subsubsection: Quasi-Poisson model}
As an alternative to NB model, Quasi-Poisson model also allows over-dispersed count data, and is a straightforward elaboration of the GLM. Instead of specifying a well-defined probability distribution for count data, Quasi-Poisson model only needs a mean model and a variance function, $\mathbb{V}(Y_{ij}) = \theta \mu_{ij}$ (with $\theta \geq 1$). While the variance-mean relationship is linear for the Quasi-Poisson model, the relationship is quadratic in NB model. This results in small foci counts being weighted more and can have greater adjustment effect in Quasi-Poisson model, which theoretically might be a perfect fit to our scenario that most brain regions have zero or low foci counts (\citealp{ver2007quasi}).

Quasi-Poisson model can be framed as GLM, with mean and variance for voxel $j$ in study $i$ is given by, 
\begin{equation}
\label{eqn: IRLS for Quasi-Poisson model}
    \begin{split}
        E[Y_{ij}] &= \mu_{ij} \\
        \mathrm{Var}(Y_{ij}) &= \theta \mu_{ij}.
    \end{split}
\end{equation}
Without a likelihood function, we instead use
ILRS algorithm, with the $(j+1)^{th}$ iteration given by, 
\begin{equation}
    \begin{split}
        \hat{\beta}^{[j+1]} &= \beta^{[j]} + (X^{*\top} W^{[j]}X^*)^{-1} X^{*\top} (Y-\mu^{[j]}) \\
        \hat{\gamma}^{[j+1]} &= \hat{\gamma}^{[j]} + (Z^{*\top} W^{[j]} Z^*)^{-1} Z^{*\top} (Y-\mu^{[j]})
    \end{split}
\end{equation}
where $W = \mathrm{diag}(\frac{\mu_{11}}{\theta}, \cdots, \frac{\mu_{1N}}{\theta}, \cdots, \frac{\mu_{M1}}{\theta}, \cdots, \frac{\mu_{MN}}{\theta})$, and $X^*=\bold{1}_M \otimes X$, $Z^*=\bold{1}_N \otimes Z$. This model can be simplified as well, though we again defer that to the next \Cref{subsection: Model factorisation}.

\subsection{Model factorisation}
\label{subsection: Model factorisation}
Having derived the explicit log-likelihood functions for meta-regression with three stochastic likelihood-based models, as well as the updating equation for a quasi-likelihood based model, we now consider model factorisation to replace the full $(MN)$-vector of foci counts by sufficient statistics. Following the generic formulation of GLM proposed in \Cref{subsubsection: Generic regression structure}, 
\begin{equation}
\label{eq:GLM formulation}
    \eta_{ij} = \log(\mu_{ij})= \sum \limits_{k=1}^P X_{jk} \beta_{k} + \sum \limits_{s=1}^R Z_{is} \gamma_s. 
\end{equation}
$\eta_{ij}$ is the estimated linear response from GLM, specific to each voxel $j$ in each individual study $i$. In this application, there are always at least $220,000$ voxels ($N$), hundreds or thousands of studies $M$, and as many as $P=456$ basis elements (with $20mm$ knots spacing), giving rise to millions of rows ($MN$) and billions of entries ($MN\times(P+R)$) in a GLM formulation. Thus, we propose a reformulation of this model into a series of sufficient statistics that are never larger than $M$ or $N$ in dimension. First, note that the localised spatial effect $\mu^X$ and global effect of study-level covariates $\mu_i^Z$ for study $i$ factorise $\mu_{ij}$ as
\begin{equation}
\mu_{ij} = \exp\left(\sum \limits_{k=1}^P X_{jk} \beta_{k} + \sum \limits_{s=1}^R Z_{is} \gamma_s\right)
= \exp\left(\sum \limits_{k=1}^P X_{jk} \beta_{k}\right) \exp\left(\sum \limits_{s=1}^R Z_{is} \gamma_s\right)
= \mu^X_{j} \, \mu^Z_i
\end{equation}
This model is identifiable as long as we ensure each covariate $Z_s$ is mean zero, letting $X$ capture the overall mean. To further simplify the total log-likelihood function, we also use the fact that $Y_{ij} \leq 1$ (either $0$ or $1$), as there will never be more than $1$ foci at the same location in a given study. Define the following notation:
\begin{itemize}
    \item Let N-vector $\mu^X = \exp(X \beta)$ be the vector of localised spatial effect of studies; 
    \item let $M$-vector $\mu^Z = \exp(Z \gamma)$ be the vector of global study-level covariates effect of studies;
    \item Let $Y_{\centerdot,j} = \sum \limits_{i=1}^M Y_{ij}$ be the sum of foci counts at voxel $j$ across all studies, and the $N-$vector $Y_{\centerdot,}=[Y_{\centerdot, 1}, \cdots, Y_{\centerdot, N}]^\top$;
    \item $Y_{i,\centerdot}= \sum \limits_{j=1}^N Y_{ij}$ be the sum of foci counts for study $i$ across all voxels, and the $M-$vector $Y_{, \centerdot}=[Y_{1, \centerdot}, \cdots, Y_{M, \centerdot}]^\top$. 
\end{itemize}
The simplified factorisation of total log-likelihood functions or IRLS updating equation are specific to each stochastic model (see \ref{appendix: Model factorisation} in Supplementary material), 
\begin{itemize}
    \item Poisson model:
    \begin{equation}
        l(\beta, \alpha) = Y_{\centerdot, }^{\top} \log(\mu^X) + Y_{,\centerdot}^{\top} \log(\mu^Z) - \left[ \boldsymbol{1}^{\top} \mu^X \right] \left[\boldsymbol{1}^{\top} \mu^Z \right],
    \end{equation} 
    \item NB model: As described in \Cref{subsubsection: Negative Binomial model}, we approximate a sum of independent NB variables again as a NB:
    \begin{equation}
        Y_{\centerdot,j} = \sum \limits_{i=1}^M Y_{ij} \sim \mathrm{NB}(r'_{j}, p'_{j}) = \mathrm{NB}( \frac{(\mu^X_j)^2  [\mathbf{1}^\top \mu^Z]^2}{\alpha' \sum \limits_{i=1}^M (\mu^X_j \mu^Z_i)^2}, \frac{\sum \limits_{i=1}^M (\mu^X_j \mu^Z_i)^2}{{(\alpha')}^{-1}\mu^X_j [\mathbf{1}^\top \mu^Z]^\top+\sum \limits_{i=1}^M (\mu^X_j \mu^Z_i)^2})
    \end{equation}
    with dispersion parameter $\alpha' = \frac{\alpha \sum \limits_{i=1}^M (\mu^X_j \mu^Z_i)^2}{(\mu^X_j)^2 [\mathbf{1}^\top \mu^Z]^2}$. The log-likelihood function is given by, 
    \begin{equation}
    l(\alpha', \beta, \gamma) = \sum \limits_{j=1}^N \left[\log\Gamma(Y_{\centerdot,j} + r'_{j}) - \log\Gamma(Y_{\centerdot,j}+1) - \log\Gamma(r'_{j}) + r'_{j} \log{(1-p'_{j}) + Y_{\centerdot,j} \log{p'_{j}}} \right],
    \end{equation}
    \item Clustered NB model:
    \begin{equation}
    \begin{split}
    l(\alpha, \beta, \gamma) &= M \alpha^{-1}\log(\alpha^{-1}) -  M \log\Gamma(\alpha^{-1}) +  \sum \limits_{i=1}^M \log\Gamma(Y_{i,\centerdot}+\alpha^{-1}) \\
    &- \sum \limits_{i=1}^M (Y_{i,\centerdot}+\alpha^{-1})\log(\alpha^{-1}+\mu_{i,\centerdot}) +  Y_{\centerdot,}^{\top} \log(\mu^X) + Y_{,\centerdot}^{\top} \log(\mu^Z)
    \end{split}
    \end{equation}
    where dispersion parameter $\alpha$ measures the excess variance across all studies and all voxels,
    \item Quasi-Poisson model: 
    \begin{equation}
    \label{eqn: IRLS Quasi-Poisson refactorisation}
    \begin{split}
        \hat{\beta}^{[j+1]} &= \beta^{[j]} + (X^\top W^{[j]}X)^{-1} X^\top (Y_{\centerdot,}-(\mu^X)^{[j]}) \\
        \hat{\gamma}^{[j+1]} &= \hat{\gamma}^{[j]} + (Z^\top V^{[j]} Z)^{-1} Z^\top (Y_{, \centerdot}-(\mu^Z)^{[j]})
    \end{split}
    \end{equation}
    where $W = \mathrm{diag}(\frac{\mu_{1}^X}{\theta}, \cdots, \frac{\mu_{N}^X}{\theta})$ and $V=\mathrm{diag}(\frac{\mu_1^Z}{\theta}, \frac{\mu_2^Z}{\theta}, \cdots, \frac{\mu_M^Z}{\theta})$.
\end{itemize}

\subsection{Fisher Scoring and optimisation via L-BFGS}
\label{subsection: Fisher scoring and optimisation}
Based on explicit log-likelihood functions associated with three stochastic likelihood-based models (Poisson, NB and clustered NB model) in \Cref{subsubsection: Poisson model}-\Cref{subsubsection: Clustered Negative Binomial model}, we employ Fisher scoring for iterative optimisation of parameters in GLMs. 
Fisher scoring replaces the gradient and Hessian of Newton's method with the score and observed Fisher's information, respectively (\citealp{longford1987fast}). Writing $\theta$ for all parameters, the updating equation at $(j+1)^{th}$ iteration is,
\begin{equation}
    \theta^{[k+1]} = \theta^{[k]} + I(\theta^{[k])})^{-1} \frac{\partial}{\partial \theta^{[k]}} l(\theta^{[k])})
\end{equation}
where the observed Fisher information is $I(\theta^{[k]})=\mathbb{E} \left[-\frac{\partial^2 l(\theta)}{\partial \theta \partial \theta^\top}\right]_{\theta=\theta^{[k]}}$.

For the Poisson model, $\theta=[\beta, \gamma]$, the Fisher information is given by,
\begin{equation}
    I(\theta) = I(\beta, \gamma) 
    = \begin{bmatrix}
    -\frac{\partial^2 l}{\partial \beta \partial \beta^\top} & -\frac{\partial^2 l}{\partial \beta \partial \gamma^\top} \\
    -\frac{\partial^2 l}{\partial \gamma \partial \beta^\top} & 
    -\frac{\partial^2 l}{\partial \gamma \partial \gamma^\top}
    \end{bmatrix}
\end{equation}
with negative Hessian matrix of $\beta$, $\left(-\frac{\partial^2 l}{\partial \beta \partial \beta^\top}\right)_{P \times P}=X^\top \mathrm{diag}(\mu^X) X^\top$; the negative cross term  $\left(-\frac{\partial^2 l}{\partial \beta \partial \gamma^\top}\right)_{P \times R} = \left(-\frac{\partial^2 l}{\partial \gamma \partial \beta^\top}\right)^\top_{R \times P}=[X^\top \mu^X][(\mu^Z)^\top Z]$; and negative Hessian matrix of $\gamma$, $\left(-\frac{\partial^2 l}{\partial \gamma \partial \gamma^\top}\right)=Z^\top \mathrm{diag}(\mu^{Z}) Z$.

Our other stochastic models with study-level covariates lead to more complicated derivations of updating equations via Fisher scoring. Instead, we use a more efficient quasi-Newton algorithm (L-BFGS algorithm, \citealp{shanno1970conditioning}), which approximates the observed Fisher information from gradient evaluations instead, and only requires limited memory and reduces the computation complexity.

\subsection{Statistical inference}
\label{subsection: Statistical inference}

\subsubsection{Global test of model fitness}
\label{subsubsection: Global test of model fitness}
Among the proposed stochastic models in \Cref{subsection: Stochastic model}, Poisson, NB and clustered NB model are likelihood-based, while Quasi-Poisson model is Quasi-likelihood based (its exact likelihood is computationally infeasible). To compare the goodness of fit from a global perspective, we will utilise likelihood-based comparison criteria (e.g., LRT, Akaike information criterion (AIC), Bayesian information criterion (BIC)) with likelihood-based models, as well as other global model fitness criteria across all stochastic models within this meta-regression framework.

\textbf{Likelihood-based model selection criteria} 
LRT uses the difference in log-likelihoods to test the null hypothesis that true model is the smaller nested model. As Poisson model is nested in both NB model and clustered NB model with dispersion parameter $\alpha=0$, for the null hypothesis $H_0$: dispersion parameter $\alpha=0$, the likelihood-ratio test statistic is given by,
\begin{equation*}
    \lambda_{LR} = -2 \left[l(\hat{\theta}_0) - l(\hat{\theta})\right]
\end{equation*}
where $l(\hat{\theta})=l(\hat{\alpha}, \hat{\beta}, \hat{\gamma})$ is maximum log-likelihood of NB model or clustered NB model without any constraint on parameters, and $l(\hat{\theta}_0)=l(\hat{\alpha}=0, \hat{\beta}, \hat{\gamma})$ is maximum log-likelihood of NB model or clustered NB model with dispersion parameter $\alpha$ constrained at $0$ (i.e. Poisson model). The test statistic is Chi-square distributed, with degree of freedom equals to $1$.

AIC and BIC are two alternatives of LRT which also deal with the trade-off between the goodness of fit and simplicity of the model, and resolve overfitting problem by penalising the number of parameters in the model. To measure the goodness of fit of a model $M$ on dataset $D$, 
\begin{equation}
    AIC = 2k-2l(\hat{\theta}), \;\;
    BIC = k \ln(n) - 2 l(\hat{\theta})
\end{equation}
where $l(\hat{\theta})$ is the maximised log-likelihood function of the model $M$, $k$ is the number of parameters in model $M$ and $n$ is the number of data points in the dataset $D$. The model with smaller AIC or BIC value is believed to be a better fit to the dataset.  

\textbf{Bias and variance of estimation}
For the purpose of selecting the best model in terms of goodness of fit on a variety of datasets, we extend the model comparisons to all stochastic models proposed in \Cref{subsection: Stochastic model}, including Quasi-Poisson model. As the central outcome of this meta-regression framework is voxelwise intensity estimation for each study, with the effect of study-level covariates being considered, it's natural to utilise bias and variance of intensity estimation as new criteria stated below, 
\begin{itemize}
    \item Relative bias of estimated total sum of intensity (per study), comparing with the averaged sum of foci count (per study) across multiple datasets;
    \item Relative bias of standard deviation (Std) in each of $x,y,z$ dimension, comparing with the actual Standard deviation in foci count (per study) across multiple datasets; 
    \item Relative bias of voxel-wise variance between actual foci count (per study) and intensity estimation (per study).
\end{itemize}
Here, relative bias is evaluated instead of bias, especially when applied to a variety of datasets with diverse foci counts.

\subsubsection{Localised inference with Wald tests on $\mu^X_{ij}$ and $\eta^X_{ij}$}
While our model is parameterised by $P$ basis elements, users want to make inference at each of the $N$ voxels. Hence, we will also explore localised inference on estimated spatial intensity $\mu^X_{ij}$ (or $\eta^X_{ij} = \log(\mu^X_{ij})$) and regression coefficient of study-level covariates ($\gamma$) via Wald tests.

\textbf{Test of spatial homogeneity: }
In the CBMA context, the most basic inference is a test of homogeneity to identify regions where more foci arise than would be expected if there were no spatial structure. Precisely, we use the null hypothesis on voxelwise intensity estimation or estimated linear response, $H_0: \mu^X_{ij} = \mu_0 = \sum \limits_{i=1}^M \sum \limits_{j=1}^N Y_{ij} / (MN)$ or $\eta^X_{ij} = \eta_0 = \log(\mu_0)$  at voxel $j$, for study $i$. The standard error for $\beta$ can be asymptotically estimated from the inverse of observed Fisher Information matrix, which gives rise to the standard error for the linear response $\eta^X_{ij}$, and thus standard error for $\mu^X_{ij}$ is obtained via delta method. It allows inference via Wald tests by examining voxelwise intensity estimation against null hypothesis of homogeneity over space. The signed Wald statistic for $\mu^X_{ij}$ or $\eta^X_{ij}$ takes the form:
\begin{equation}
    Z_{\mu^X} = \frac{\mu^X_{ij}-\mu_0}{\mathrm{SE}(\mu^X_{ij})}, \; \;
    Z_{\eta^X} = \frac{\eta^X_{ij}-\eta_0}{\mathrm{SE}(\eta^X_{ij})}
\end{equation}
where $SE(\mu^X_{ij})$ is the standard error of estimated spatial intensity $\mu^X_{ij}$, and $SE(\eta^X_{ij})$ is the standard error of estimated linear response $\eta^X_{ij}$, and the statistics are Gaussian asymptotically. Finally, we can create p-value maps that are thresholded to control FDR at $5\%$ (\citealp{benjamini1995controlling}). 

\subsubsection{Inference on study-level covariates}
\label{subsubsection: Inference on study-level covariates}
For regression coefficient $\gamma$ ($s \times 1$) of study-level covariates, we consider general linear hypothesis (GLH) tests through a contrast matrix $C_\gamma$ ($m \times s$). Under the null hypothesis,
\begin{equation}
\label{equation: study-level covariates H_0}
    H_0: C_{\gamma} \gamma = \bold{0}_{m \times 1}
\end{equation}
The test statistic follows a $\chi^2$ distribution with $m$ degree of freedom asymtotically,
\begin{equation}
    (C_{\gamma} \hat{\gamma})^T (C_\gamma \mathrm{Cov}(\hat{\gamma}) C^T_\gamma)^{-1} (C_\gamma \hat{\gamma}) \xrightarrow{D} \chi^2_m
\end{equation}
and in the case of a single contrast $(m=1)$, a signed Z test can be computed. Details of GLH on study-level covariates can be found in \ref{Appendix: Contrasts on regression coefficient of study-level covariates} in Supplementary material.

\section{Experiments} 
\label{section: Experiments}
\subsection{Simulation settings}
The statistical analyses of model estimation with CBMA data are conducted at voxel level: voxelwise test statistics are evaluated to examine the significance of experimental effect. Therefore, before investigating the model fitness, we evaluate our models' false positive rates (FPR) under null settings. Due to the computationally intensive nature of these evaluations, we only evaluated the two models that showed promise in other evaluations, Poisson and NB. Under the null hypothesis of spatial homogeneity, we use Monte Carlo (MC) simulation to establish the validity of FPR control for the test of spatial intensity ($\mu^X$). Specifically, we will explore meta-regression with Poisson or NB model, either with non-null study-level covariates or without study-level covariates. To ensure the validity of FPR control is applicable to all CBMA data, the sampling mechanism is either model-based or empirical, with simulated foci count always analogous to the foci count within a real dataset. Specifically, in model-based sampling, data generating mechanism matches with regression model, with number of studies and average foci per study identical to the original dataset; while in empirical sampling, real data foci locations are randomly shuffled to guarantee the spatial homogeneity of foci distribution.

\subsection{Applications to $20$ meta-analytic datasets}
\label{subsection: Applications to }
Cognition concerns psychological and cognitive processes that focus on learning people's perception, interpretation and response to information and stimuli. It refers to both conscious procedure and unconscious, automatic mechanisms in the brain that occur as a response to stimuli, and is highly variable across individuals (\citealp{gallagher2019neurocognitive}). Cognition has been studied intensively to identify brain regions that are involved in cognition tasks, conducted in an MRI scanner. For the purpose of evaluating the accuracy and sensitivity of this meta-regression framework, as well as analysing goodness of fit of stochastic models with respect to different CBMA datasets, $20$ previously published meta-analytic datasets are used in this article, which involves multiple aspects of cognition research, as well as other stimulus-based and diagnosis-based research (as displayed in \Cref{table:20 meta-analytic datasets}). 

\begin{table*}[!htb]
  \centering
  \footnotesize
  \begin{threeparttable}[]
  \caption{Number of contrasts and foci counts of $20$ meta-analytic datasets}
  \label{table:20 meta-analytic datasets}
  \begin{tabular}{l|l|l|l|l}
    \toprule
    Dataset  & number of contrasts & total count of foci & max foci count & average foci count\\
    \midrule
    1. Social Processing & $599$ & $4934$ & $47$ & $8.24$ \\
    2. PTSD & $22$ & $154$ & $26$ & $7.00$ \\
    3. Substance Use & $89$ & $657$ & $110$ & $7.38$
    \\
    4. Dementia & $28$ & $1194$ & $548$ & $42.64$\\
    5. Cue Reactivity & $275$ & $3197$ & $58$ & $11.63$\\
    6. Emotion Regulation & $338$ & $3543$ & $87$ & $10.48$\\
    7. Decision Making & $145$ & $1225$ & $49$ & $8.45$\\
    8. Reward & $850$ & $6791$ & $59$ & $7.99$\\
    9. Sleep Deprivation & $44$ & $454$ & $59$ & $10.32$\\
    10. Naturalistic & $122$ & $1220$ & $59$ & $10.00$\\
    11. Problem Solving & $282$ & $3043$ & $44$ & $10.79$\\
    12. Emotion & $1738$ & $22038$ & $203$ & $12.68$\\
    13. Cannabis Use & $81$ & $314$ & $16$ & $3.88$\\
    14. Nicotine Use & $13$ & $77$ & $23$ & $5.92$\\
    15. Frontal Pole CBP & $795$ & $9525$ & $57$ & $11.98$\\
    16. Face Perception & $385$ & $2920$ & $50$ & $7.58$\\
    17. Nicotine Administration & $75$ & $349$ & $24$ & $4.65$\\
    18. Executive Function & $243$ & $2629$ & $54$ & $10.82$\\
    19. Finger Tapping & $76$ & $696$ & $27$ & $9.16$\\
    20. n-Back & $29$ & $640$ & $69$ & $22.07$\\
    \bottomrule
  \end{tabular}
  \label{table: p-value for LRT}
  \end{threeparttable}
\end{table*}

\begin{figure}[h]
\includegraphics[width=14cm]{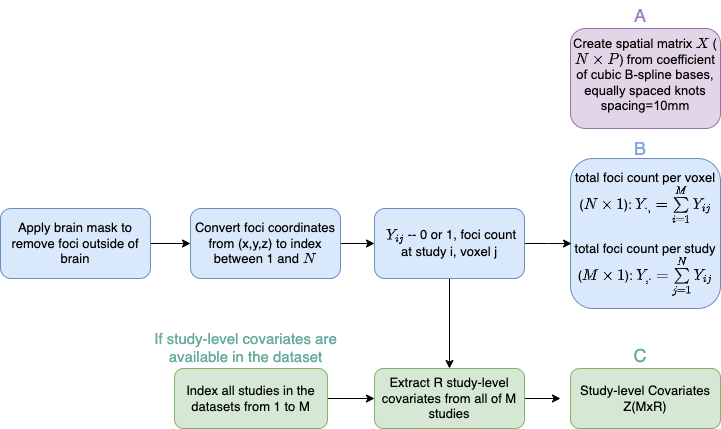}
\centering
\caption{Preprocessing pipeline of $20$ meta-analytic datasets before fitting CBMR framework. Note that panel A and B are applicable to all datasets, which generate a spatial design matrix $X$, total foci count per voxel $Y_{\centerdot,}(N \times 1)$ and total foci count per study $Y_{, \centerdot} (M \times 1)$. While covariates matrix $Z(M \times R)$ in panel C is integrated into CBMR only if the effect of study-level covariates is considered.} 
\label{figure: preprocess_pipeline}
\end{figure}

Preprocessing steps are summarised in \Cref{figure: preprocess_pipeline}. The discrete sampling space of our analysis is the 2$mm^3$ MNI atlas, with dimensions $91\times109\times91$, and $N=228483$ brain voxels. We first apply this brain mask to remove foci outside the brain and remove any multiple-foci (while original data peaks are always distinct, a foci count in excess of 1 cannot occur after rounding to the 2mm space). We then extract all the sufficient statistics after model factorisation in \Cref{subsection: Model factorisation}, including spatial design matrix $X (N \times P)$ generated from B-spline bases, total foci count per voxel $Y_{\centerdot ,} (N \times 1)$ and total foci count per study $Y_{, \centerdot}(M \times 1)$ and study-level covariates $Z(M \times R)$ if considered. 

\section{Results}
\label{section: Results}
\subsection{Simulation results}
\label{subsection: Simulation results}
For each of the $20$ meta-analytic datasets, we simulate foci distribution under a null hypothesis of spatial homogeneity, estimate spatial intensity and investigate the distribution of voxel-wise p-values for the eight different scenarios: fitting Poisson or NB model, use of a model-based or empirical (random shuffling) data sampling mechanism, and use or omission of study-level covariates; for all settings we use B-spline knot of spacing $20mm$ in $x,y,z$ direction, producing $P=456$ basis elements. The computation of test statistics depends on the covariance of regression coefficients, which is approximated by the inverse of Fisher Information matrix of optimised parameters at maximised log-likelihood (see \Cref{subsection: Fisher scoring and optimisation}). Empirically, we sometimes observe the Fisher Information matrix is ill-conditioned and numerically singular, which is associated with datasets having low foci count. Especially in datasets with few studies, some regions have essentially no foci, leading some of the $\beta$ coefficients to be driven to negative infinity and produce an estimated rate of zero. In the experiments, we found that datasets with a total foci count of at least 200 generally avoided these singularity problems and produced accurate standard errors.

\begin{figure}[h]
\centering
\includegraphics[width=14cm]{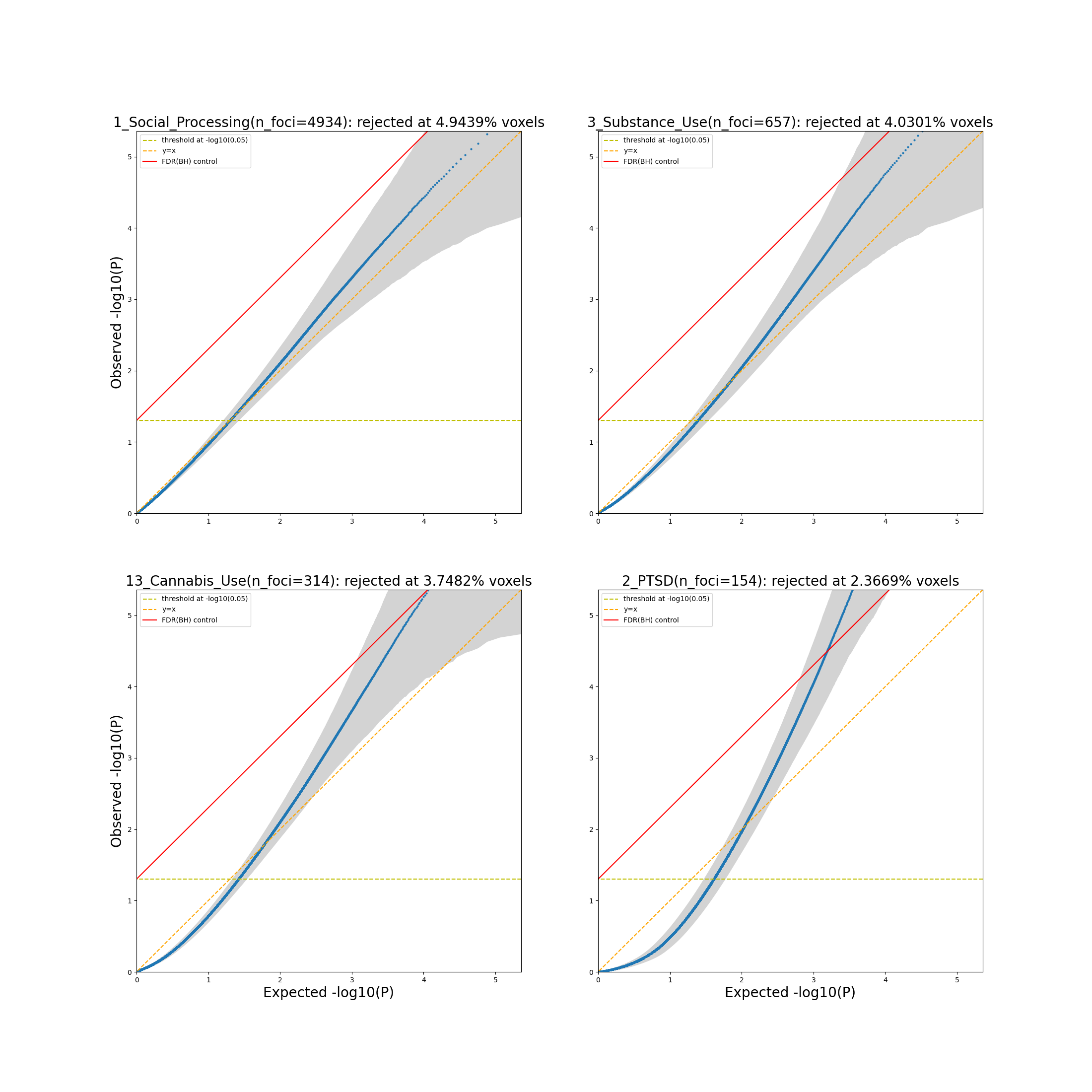}
\caption{P-P plot of $p$-value (under $-\log_{10}$ scale) with four representative meta-analytic datasets (Social Processing, Substance Use, Cannabis Use and PTSD datasets), estimated by CBMR with Poisson model without study-level covariates, sampled with model-based approach.} 
\label{figure: representative PP-plots}
\end{figure}

To establish the validity of spatial homogeneity tests ($\mu_{j}^X=\mu_0$, $\forall j=1, \cdots, N$) for each $20$ meta-analytic datasets, we compute p-values and create P-P plots. We compute 100 null realisations, each producing N p-values (one for each voxel), with the null expected $-\log_{10}$ p-values ranging from $-\log_{10}(N/(N+1))\approx0$ to $-\log_{10}(1/(N+1))=5.359$. For each expected null p-value, we plot the mean and 95$\%$ prediction interval via a normal approximation (mean $\pm$ 1.96 $\times$ standard deviation, computed over 100 realisations). Since the P-P plots are very similar for each of the eight scenarios, we only display results for the setting of CBMR with Poisson model without study-level covariates, sampled with a model-based approach. \Cref{figure: representative PP-plots} shows the four representative $-\log_{10}$ P-P plots (results for all 20 studies shown in \Cref{figure: full PP-plots} in Supplementary material), with identity (dashed diagonal line), $5\%$ significance (dashed horizontal line) and the FDR $5\%$ boundary (solid diagonal line); gray shaded areas plot the point-wise $95\%$ prediction intervals. It shows that p-values $< 0.05 \approx 10^{-1.3}$ are valid, and extreme p-values can skew liberal; the worst affected cases are datasets with very few foci (e.g. analysis 14).  In general, datasets with total foci counts less than 200 show poor behaviour.

Since multiple testing correction requires valid p-values far smaller than $0.05$, we focus on control of FDR in these null simulations.  None of the 20 datasets have valid FDR control (PP-plots or prediction intervals fall above the $5\%$ Benjamini-Hochberg threshold). However, the PP plots generally show valid p-values $< 10^{-3}$, and if we truncate p-values by replacing any p-value smaller than $10^{-3}$ with that value, we obtain valid (if conservative) FDR control  (\Cref{table: percentage of FDR control}). This pragmatic approach could impact power, but empirical results (\Cref{subsection: Results from $20$ meta-analytic datasets}) suggest the inferences based on truncated p-values remain sensitive.

\begin{table*}[!htb]
  \centering
  \footnotesize
  \begin{threeparttable}[]
  \caption{The percentage of invalid FDR control (before/after p-value truncated at $10^{-3}$) in $20$ meta-analytic datasets over $100$ realisations}
  \label{table:few_shot}
  \begin{tabular}{l|l|l|l|l|l}
    \toprule
    Dataset  & Before & After & Dataset & Before & After \\
    \midrule
    1. Social Processing & $44\%$ & $0\%$ & 2. PTSD & $100\%$ & $0\%$ \\
    3. Substance Use & $26\%$ & $0\%$ & 4. Dementia & $16\%$  & $0\%$ \\
    5. Cue Reactivity & $28\%$ & $0\%$ & 6. Emotion Regulation & $23\%$ & $0\%$ \\
    7. Decision Making & $18\%$ & $0\%$ & 8. Reward & $43\%$ & $0\%$ \\
    9. Sleep Deprivation & $30\%$ & $0\%$ & 10. Naturalistic & $22\%$ & $0\%$ \\
    11. Problem Solving & $26\%$ & $0\%$ & 12. Emotion & $100\%$ & $0\%$ \\
    13. Cannabis Use & $63\%$ & $0\%$ & 14. Nicotine Use & $94\%$ & $0\%$ \\
    15. Frontal Pole CBP & $90\%$ & $0\%$ & 16. Face Perception & $19\%$ & $0\%$ \\
    17. Nicotine Administration & $54\%$ & $0\%$ & 18. Executive Function & $22\%$ & $0\%$ \\
    19. Finger Tapping & $22\%$ & $0\%$ & 20. n-Back & $27\%$ & $0\%$ \\
    \bottomrule
  \end{tabular}
  \label{table: percentage of FDR control}
  \end{threeparttable}
\end{table*}

\subsection{Results from $20$ meta-analytic datasets}
\label{subsection: Results from $20$ meta-analytic datasets}
We first evaluate the goodness of fit among likelihood-based stochastic models (Poisson, NB and clustered NB model) via comparisons of maximised log-likelihood, AIC and BIC. As shown in \Cref{figure: log_likelihood plot} in \ref{Appendix: Likelihood-based comparison} of Supplementary material, CBMR with NB model outperforms two other likelihood-based stochastic models in every dataset. Not surprisingly, as NB model is the only stochastic approach that allows for anticipated excess variance relative to Poisson at voxel level; clustered NB is better than Poisson for the majority of these $20$ meta-analytic datasets, but only by a small margin. It is conceivable that although a study-wise global dispersion parameter exists in clustered NB model, CBMA data is completely specified by a Poisson model at voxel level. As there exists nested relationships between Poisson and both NB and clustered NB model (with dispersion parameter $\alpha=0$), we also conduct LRT to evaluate the trade-off between sufficiency and complexity of model. We found the null hypothesis that the nested model (Poisson) is better than full model (NB) is rejected for all datasets, with $p$-value less than $10^{-8}$. The clustered NB model is preferred over Poisson for the majority of the $20$ meta-analytic datasets (with $p$-value less than $10^{-8}$) (more details to follow in \Cref{table: p-value for LRT} in \Cref{Appendix: Likelihood-based comparison} of Supplementary material).

\begin{figure}
\begin{minipage}{.5\linewidth}
\centering
\subfloat[]{\label{main:a}\includegraphics[width=8cm]{figures/Bias_foci_sum_plot.pdf}}
\end{minipage}
\begin{minipage}{.5\linewidth}
\centering
\subfloat[]{\label{main:b}\includegraphics[width=8cm]{figures/Bias_std_plot.pdf}}
\end{minipage}\par\medskip
\centering
\subfloat[]{\label{main:c}\includegraphics[width=8cm]{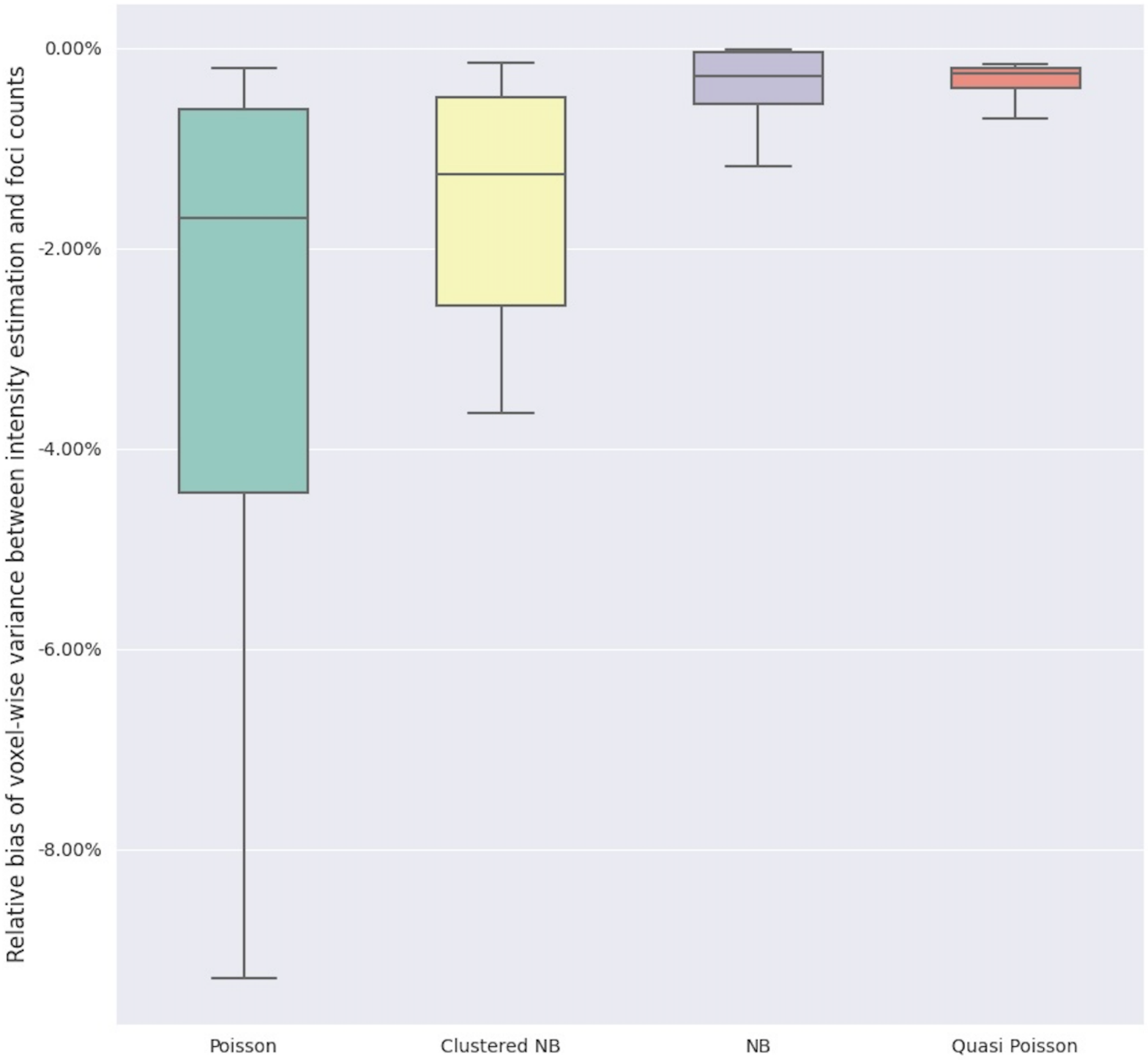}}
\caption{Results from bias-related model comparison criteria, fitted with four stochastic models on each of $20$ meta-analytic datasets: (a) Boxplots of relative bias of sum of intensity estimation (per study); (b) Boxplots (each of $x,y,z$ dimensions) of relative bias of standard deviation of intensity estimation (per study); (c) Boxplots of relative bias of voxelwise variance of intensity estimation (per study).}
\label{fig: Bias and var of model estimation}
\end{figure}

Apart from model comparisons via likelihood-based criteria (LRT, AIC and BIC), we also integrate Quasi-Poisson model into comparisons of model fitness, based on bias and variance criteria which are applicable to Quasi-likelihood models. The results in \Cref{fig: Bias and var of model estimation}(a) suggest that four stochastic models (Poisson, NB, clustered NB and Quasi-Poisson model) all give rise to accurate intensity estimation with relative bias of sum of intensity estimation (per study) less than $0.5\%$, among which Poisson model has the lowest median relative bias ($0.05\%$) across $20$ meta-analytic datasets. Quasi-Poisson and NB model display both underestimation and overestimation of study-wise sum of intensity across $20$ meta-analytic datasets, while sum of intensity are always overestimated with Poisson and clustered NB model. The results in \Cref{fig: Bias and var of model estimation}(b) suggest that CBMR framework also provides an accurate estimation of standard variation (Std) of intensity in each of $x,y,z$ dimension, with relative
bias controlled within $-0.2\%$ to $0.1\%$ for all stochastic models across $20$ meta-analytic datasets, and estimated intensity along $x$ axis are the most accurate (with smallest Std bias). As shown in \Cref{fig: Bias and var of model estimation}(c), CBMR with Poisson model displays the largest negative bias of variance between intensity estimation and foci count, which suggests that excess variance cannot be explained by Poisson assumption (only voxels with nonzero foci count are studied and included in this plot). Clustered NB model displays second largest negative relative bias in voxel-wise variance estimation (per study), which is potentially related to the fact that it evaluates a study-specific over-dispersion parameter over space, but the intensity function is modelled by Poisson model at voxel level. Small relative bias is found in both NB and Quasi-Poisson model (with median $-0.28\%$ and $-0.25\%$), with less variation in relative bias across multiple datasets with Quasi-Poisson model, which suggests both models are capable of dealing with excess variance in CBMA data. 

Overall, we regard these evaluations as an evidence that NB model is preferred. While it has slight bias for total intensity (\Cref{fig: Bias and var of model estimation}(c)), it has much more accurate variance than the Poisson model.

\subsection{Comparison with ALE}
\label{subsection: Comparison with ALE}
\begin{figure}
\centering
\begin{subfigure}[b]{1\textwidth}
   \includegraphics[width=14cm]{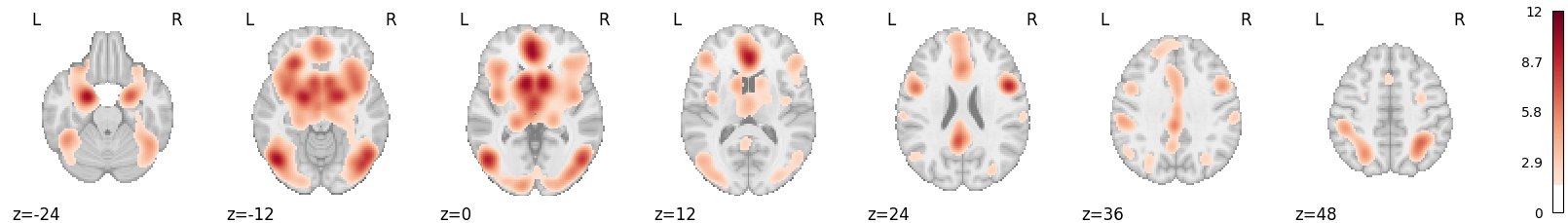}
   \caption{Z-score map generated by ALE}
   \label{fig:ALE_comparison_a} 
\end{subfigure}

\begin{subfigure}[b]{1\textwidth}
   \includegraphics[width=14cm]{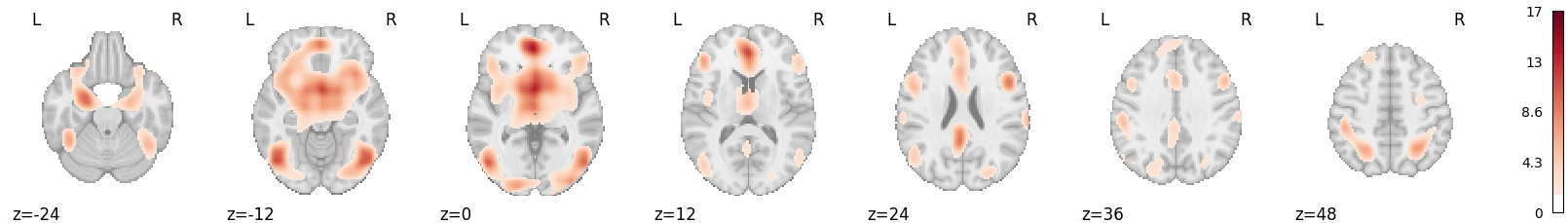}
   \caption{Z-score map generated by CBMR (with Poisson)}
   \label{fig:ALE_comparison_b}
\end{subfigure}

\begin{subfigure}[c]{1\textwidth}
   \includegraphics[width=14cm]{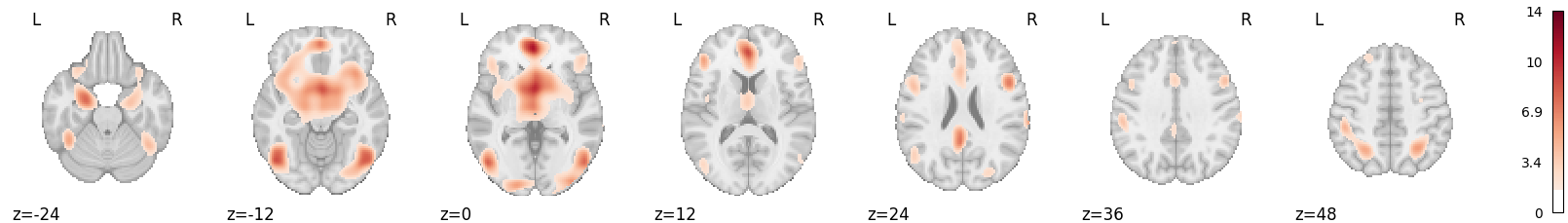}
   \caption{Z-score map generated by CBMR (with NB)}
   \label{fig:ALE_comparison_c}
\end{subfigure}
\begin{subfigure}[c]{1\textwidth}
   \includegraphics[width=14cm]{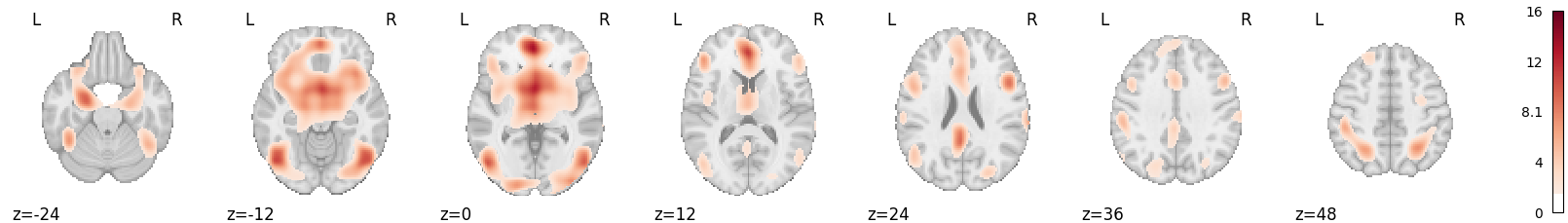}
   \caption{Z-score map generated by CBMR (with Clustered NB)}
   \label{fig:ALE_comparison_d}
\end{subfigure}
\caption{Activation maps (for significant uncorrected p-values under $5\%$ significance level, presented in Z-scores) generated by ALE (with FWHM=$14$) and CBMR (with a variety of stochastic models) on the Cue Reactivity dataset with axial slices at $z=-24,-12,0,12,24,36, 48$. Under the null hypothesis of spatial homogeneity, activation regions with z-scores corresponding to uncorrected p-values below the significance level $0.05$ are highlighted.}
\label{fig: comparison between ALE and CBMR}
\end{figure}

We compared our CBMR results to a widely used approach, computing tests for spatial homogeneity across space with both CBMA and ALE. For simplicity, we only demonstrate the comparison of detected activation regions on the Cue Reactivity dataset (total foci count of $6288$) (\citealp{hill2021cue}). For comparison purposes, we show z statistic values at all voxels significant at $\alpha=0.05$ uncorrected in \Cref{fig: comparison between ALE and CBMR}. Here, we choose FWHM=$14$ to obtain comparative spatial resolution between ALE and CBMR. Consistency in activation regions is found in left cerebral cortex, frontal orbital cortex, insular cortex, left and right accumbens, while spatial specificity of activation regions differ slightly in ALE and CBMR, with ALE detecting slightly more voxels.

 Another criterion of consistency is the dice similarity coefficient (DSC), the intersection of ALE and CMBR significant voxels divided by the average number of significant voxels. As shown in \Cref{table: DSC between ALE and CBMR}, ALE appears generally more sensitive than CBMR regardless of foci counts in the datasets, though DSC varies above $71.89\%$ to $80.33\%$ on the datasets with more than $1200$ foci counts, which demonstrates good similarity between the methods. 

\begin{table*}[!htb]
  \centering
  \footnotesize
  \begin{threeparttable}[]
  \caption{Number of voxels in activation regions of ALE (FWHM=$14$) and CBMR (with Poisson), based on uncorrected p-values with $5\%$ significance level, as well as Dice similarity coefficient in $20$ meta-analytic datasets (Datasets are listed in an ascending order according to total number of foci).}
  \begin{tabular}{l|l|l|l|l|l}
    \toprule
    Dataset  & n\_foci & $\vert AR_{CBMR} \vert $ & $\vert AR_{ALE} \vert$ & $\vert AR_{CBMR} \cap AR_{ALE} \vert$ & DSC \\
    \midrule
    14. Nicotine Use & $77$ & $1312$ & $12431$ & $1154$ & $17.79\%$ \\
    2.\;\; PTSD & $154$ & $6306$ & $15866$ & $5067$ & $45.71\%$ \\
    13. Cannabis Use & $314$ & $11841$ & $18390$ & $8235$ & $54.48\%$ \\
    17. Nicotine Administration & $349$ & $11546$ & $18916$ & $8028$ & $52.71\%$ \\
    9.\;\; Sleep Deprivation & $454$ & $10250$ & $15461$ & $5732$ & $44.59\%$ \\
    20. n-Back & $640$ & $19404$ & $31512$ & $17627$ & $69.24\%$ \\
    3.\;\; Substance Use & $657$ & $19024$ & $26477$ & $13602$ & $59.79\%$ \\
    19. Finger Tapping & $696$ & $19067$ & $33914$ & $17939$ & $67.72\%$ \\
    4.\;\; Dementia & $1194$ & $16244$ & $30437$ & $12464$ & $53.41\%$ \\
    10. Naturalistic & $1220$ & $22328$ & $29442$ & $15344$ & $59.28\%$ \\
    7.\;\; Decision Making & $1225$ & $28284$ & $36735$ & $23372$ & $71.89\%$ \\
    12. Emotion & $2038$ & $57698$ & $67699$ & $48847$ & $77.91\%$ \\
    18. Executive Function & $2629$ & $33848$ & $46679$ & $31698$ & $78.73\%$ \\
    16. Face Perception & $2920$ & $41682$ & $53109$ & $36710$ & $77.45\%$ \\
    11. Problem Solving & $3043$ & $38466$ & $51315$ & $34757$ & $77.43\%$ \\
    5.\;\; Cue Reactivity & $3197$ & $41242$ & $52371$ & $37301$ & $79.69\%$ \\
    6.\;\; Emotion Regulation & $3543$ & $36602$ & $48157$ & $31176$ & $73.56\%$ \\
    1.\;\; Social Processing & $4934$ & $48376$ & $61136$ & $40740$ & $74.40\%$ \\
    15. Frontal Pole CBP & $9525$ & $53165$ & $65339$ & $47595$ & $80.33\%$ \\
    8.\;\; Reward & $6791$ & $43048$ & $51721$ & $37711$ & $79.59\%$ \\
    \bottomrule
  \end{tabular}
  \label{table: DSC between ALE and CBMR}
  \end{threeparttable}
\end{table*}

\begin{figure}
\centering
\begin{subfigure}[b]{1\textwidth}
   \includegraphics[width=14cm]{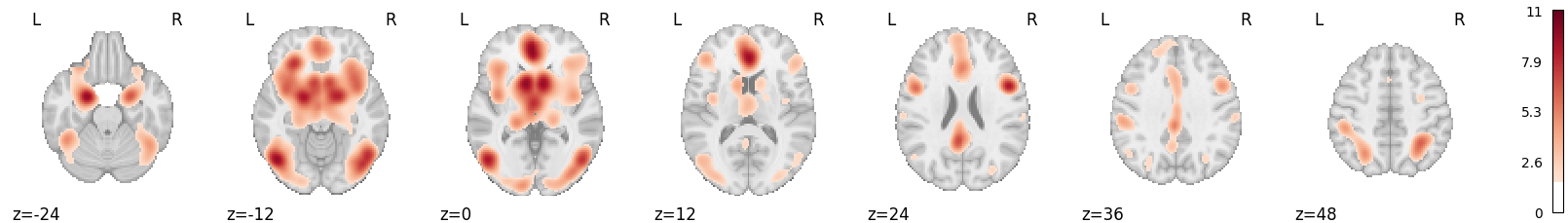}
   \caption{Z-score map generated by ALE}
   \label{fig:corrected_ALE_comparison_a} 
\end{subfigure}

\begin{subfigure}[b]{1\textwidth}
   \includegraphics[width=14cm]{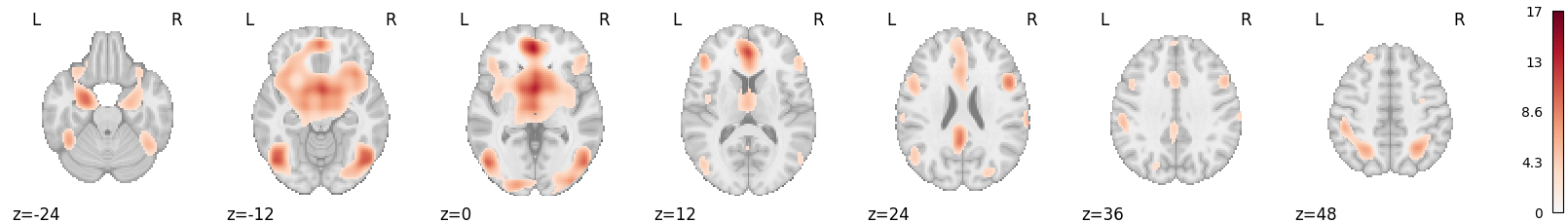}
   \caption{Z-score map generated by CBMR (with Poisson)}
   \label{fig:corrected_ALE_comparison_b}
\end{subfigure}

\begin{subfigure}[c]{1\textwidth}
   \includegraphics[width=14cm]{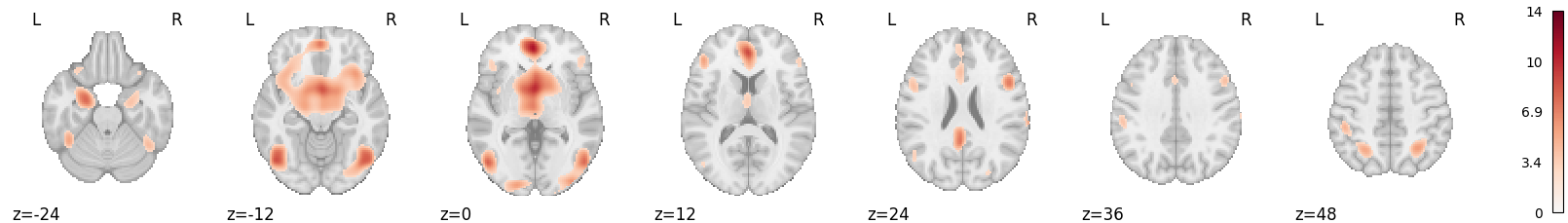}
   \caption{Z-score generated by CBMR (with NB)}
   \label{fig:corrected_ALE_comparison_c}
\end{subfigure}
\begin{subfigure}[c]{1\textwidth}
   \includegraphics[width=14cm]{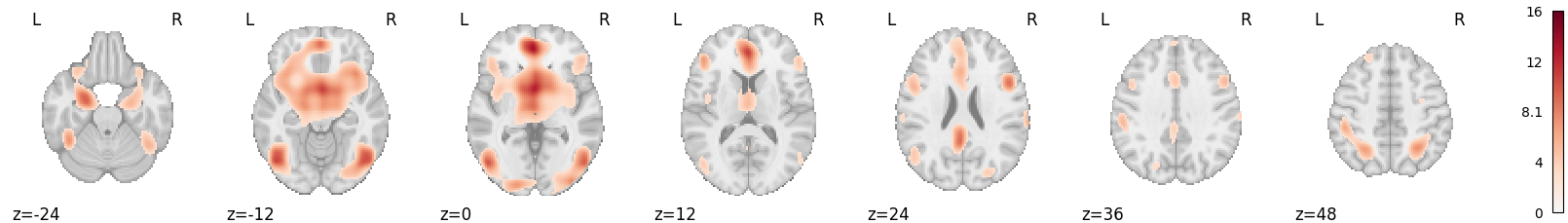}
   \caption{Z-score map generated by CBMR (with Clustered NB)}
   \label{fig:corrected_ALE_comparison_d}
\end{subfigure}
\caption{Activation maps (for significant FDR corrected p-values under $5\%$ significance level, presented in Z-scores) generated by ALE and CBMR with FDR correction (by BH procedure) with truncated p-values of Cue Reactivity dataset. The figure is shown with axial slices at $z=-24,-12,0,12,24,36, 48$. Under the null hypothesis of spatial homogeneity, activation regions with z-scores corresponding to corrected p-values below the significance level $0.05$ are highlighted.}
\label{fig: corrected comparison between ALE and CBMR}
\end{figure}

ALE evaluates experimental effect by testing probabilistic maps (generated by Gaussian kernel) against the null hypothesis, CBMR estimates activation intensity and conducts hypothesis testing at voxel level, both of them neglect the effects of testing all voxels simultaneously, and cannot control the rate of false rejections. Some researchers proposed a conservative threshold ($\alpha=0.0001$) on the uncorrected p-values to reduce type I error (\citealp{turkeltaub2002meta}), while a more principled approach is to control the false discovery rate (FDR) via Benjamini-Hochberg (BH) procedure. \Cref{fig: corrected comparison between ALE and CBMR} shows a comparison of results using a $5\%$ FDR threshold, where CMBR (Poisson) p-values use a $10^{-3}$ truncation, and \Cref{table: corrected DSC between ALE and CBMR} shows a comparison of number of detected voxels. The sensitivity of ALE and CBMR is comparable, with sometimes ALE or CBMR being detecting more voxels. The DSC varies between $70.55\%$ and $79.76\%$ for datasets with more than $1225$ foci counts, indicating consistency of activation regions between ALE and CBMR approach after FDR correction. 

\begin{table*}[!htb]
  \centering
  \footnotesize
  \begin{threeparttable}[]
  \caption{Number of voxels in activation regions of ALE (FWHM=$14$) and CBMR (with Poisson), based on FDR corrected p-values (using BH procedure) with $5\%$ significance level , as well as Dice similarity coefficient in $20$ meta-analytic datasets (Datasets are listed in an ascending order according to total number of foci).}
  \begin{tabular}{l|l|l|l|l|l}
    \toprule
    Dataset  & n\_foci & $\vert AR_{CBMR} \vert $ & $\vert AR_{ALE} \vert$ & $\vert AR_{CBMR} \cap AR_{ALE} \vert$ & DSC \\
    \midrule
    14. Nicotine Use & $77$ & $209$ & $0$ & $0$ & $0.00\%$ \\
    2.\;\; PTSD & $154$ & $0$ & $1201$ & $0$ & $0.00\%$ \\
    13. Cannabis Use & $314$ & $313$ & $152$ & $17$ & $7.31\%$ \\
    17. Nicotine Administration & $349$ & $1338$ & $943$ & $522$ & $45.77\%$ \\
    9.\;\; Sleep Deprivation & $454$ & $176$ & $0$ & $0$ & $0.00\%$ \\
    20. n-Back & $640$ & $11456$ & $17725$ & $10212$ & $69.99\%$ \\
    3.\;\; Substance Use & $657$ & $3145$ & $2082$ & $1225$ & $46.87\%$ \\
    19. Finger Tapping & $696$ & $12410$ & $23837$ & $11590$ & $63.95\%$ \\
    4.\;\; Dementia & $1194$ & $5126$ & $7931$ & $3142$ & $48.13\%$ \\
    10. Naturalistic & $1220$ & $4192$ & $3241$ & $1861$ & $50.07\%$ \\
    7.\;\; Decision Making & $1225$ & $15331$ & $20468$ & $12628$ & $70.55\%$ \\
    18. Executive Function & $2629$ & $26039$ & $37797$ & $24690$ & $77.67\%$ \\
    16. Face Perception & $2920$ & $28893$ & $38193$ & $25533$ & $76.12\%$ \\
    11. Problem Solving & $3043$ & $28221$ & $39091$ & $25675$ & $76.29\%$ \\
    5.\;\; Cue Reactivity & $3197$ & $30382$ & $38847$ & $27375$ & $78.57\%$ \\
    6.\;\; Emotion Regulation & $3543$ & $23388$ & $31620$ & $20056$ & $72.92\%$ \\
    1.\;\; Social Processing & $4943$ & $34317$ & $45263$ & $28555$ & $71.76\%$ \\
    8.\;\; Reward & $6791$ & $33021$ & $39743$ & $28728$ & $78.96\%$ \\
    15. Frontal Pole CBP & $9525$ & $44030$ & $55251$ & $39594$ & $79.76\%$ \\
    12. Emotion & $22038$ & $50480$ & $57321$ & $41918$ & $77.77\%$ \\
    \bottomrule
  \end{tabular}
  \label{table: corrected DSC between ALE and CBMR}
  \end{threeparttable}
\end{table*}

\subsection{Effect of study-level covariates}
\label{subsection: Effect of study-level covariates}
Unlike ALE, CBMR is based on explicit probabilistic models, and can estimate the effect of study-level covariates. Here, we integrate two study-level covariates, study-wise (square root) sample size and year of publication (after centring and standardisation) into CBMR framework on each of $20$ meta-analytic datasets. We find, for example, on Cue Reactivity dataset, the year of publication is not significant ($Z=-0.6880, p=0.4915$), while sample size is significant ($Z=6.1454, p<10^{-8}$) (See \Cref{table: Hypothesis testing on effect of study-level covariates} for $p-$value and $Z-$score of study-level covariates on each of $20$ meta-analytic datasets).

\section{Discussion}
\label{section: Discussion}
In this work we have presented a meta-regression framework with a spatial model as a general approach for CBMA data, where we have considered multiple stochastic models and allowed for study-level factors (e.g., sample size and year of publication). Our approach uses a spline parameterization to model the smooth spatial distribution of activation foci, and fits a generalised linear model with different variants of voxelwise (Poisson model, NB model and Quasi-Poisson model) or study-wise (Clustered NB model) statistical distributions. Our approach is a computationally efficient alternative to previous Bayesian spatial regression models, providing the flexibility and interpretability of a regression model while jointly modelling all of space. For comparison, the implementation of Bayesian log-Gaussian Cox process regression needed approximately $30$ hours on an NVIDIA Tesla K20c GPU card \cite{samartsidis2019bayesian}, while our meta-regression runs for roughly $20$ minutes on an NVIDIA GTX 1080 Graphics Card. Furthermore, as a more intuitive and interpretable approach derived from generalised linear model, we believe that our meta-regression framework is more comprehensible to practitioners, relative to the spatial posterior intensity function. Through simulations on synthetic data (with simulated foci counts analogous to foci counts in each of $20$ meta-analytic datasets), we demonstrated valid FDR control for spatial homogeneity null hypothesis after a truncation of $p-$values below $10^{-3}$. According to $20$ meta-analytic datasets, we found that NB model is the most accurate stochastic model in model comparisons via LRT, AIC and BIC, as well as smallest relative bias in both mean and variance of intensity estimation (per study), while Poisson and clustered NB model cannot explain over-dispersion observed in foci count. Meanwhile, we also compare the findings of activation regions from both ALE and CBMR approach, and justify the validity and robustness of CBMR, especially on the datasets with relatively high foci count, e.g., datasets with at least $200$ total foci. 

There are a few limitations in our work. Here we have only considered a single group of studies. In future work, we will extend our method to estimate the spatial intensity function of multiple groups (e.g., multiple types of stimuli within a cognitive task), so that we can investigate the consistency and difference of activation regions by group comparison. Meanwhile, we are currently not using regularisation term on spatial regression coefficients of CBMR. We have considered Firth-type penalty which indeed guarantees convergent estimates (especially on brain regions without any foci) and removes the first-order asymptotic bias term of maximum likelihood estimates, but it also causes a significant over-estimation of intensity at the edge of brain mask. The edge effect induced by Firth-type penalty relates to the structure of the Jeffreys prior and higher intensity associated with edge and corner basis elements. However, it's plausible to consider regularising likelihood functions with alternative penalty term (e.g., $L_1$ or $L_2$ norm) in the future, and figure out the optimal value of hyper-parameter. To estimate the variance of voxelwise spatial intensity, we approximate the covariance of spatial regression coefficients by inverting the Fisher Information matrix. It might give rise to numerical instability because the dimension of Fisher Information matrix is large (there are hundreds or even thousands elements in spline bases), and it might even be numerically singular for datasets with low foci count since most of voxels have near-zero intensity estimation. We have tried many approaches to improve numerical stability, including adding a extremely small epsilon ($10^{-6}$) or $1\%$ of the largest diagonal element on the diagonal of Fisher Information matrix, or computing the Fisher Information assuming the null hypothesis of homogeneity is true. However, all of these efforts produced underestimation of the variance of voxelwise spatial intensity and gave rise to invalid p-values. In future work, we might consider non-parametric methods to estimate the covariance of spatial regression coefficient instead of the inverse of Fisher Information, or add a regularisation term on B-spline roughness to avoid very negative spatial regression coefficients. 

Another potential direction is to conduct meta-analysis with data from multiple source, specifically, integrate additional information about reported foci or full statistic map (e.g., p-values or t-scores) if available. Some researches proposed Markov melding as a fully Bayesian framework for joining probabilistic sub-models, where evidence from different source is specified in each sub-model, and sub-models are joined while preserving all information and uncertainty (\citealp{goudie2019joining}). Such an approach might enrich the inference obtained from CBMR by integrating the magnitude of CBMA activation or even image-based meta-analysis data. Finally. it's worth considering a zero-inflated stochastic model (e.g., Poisson or NB model) as the current datasets only consist of studies with at least one focus, there might be inflated zero foci count than observed. Excess zeros are separated and modelled independently in zero-inflated models, which might provide a more accurate approximation for low-rate Binomial data. 

\section*{Software}
\label{section: Software}
Implementation in the form of Python and Pytorch code can be found in \href{https://github.com/yifan0330/CBMR}{Github repository}. CBMR framework has also been implemented and integrated into \href{https://nimare.readthedocs.io/en/latest/index.html#}{NiMARE python package}.

\section*{Acknowledgments}
The computational aspects of this research were supported by the Wellcome
Trust Core Award Grant Number 203141/Z/16/Z and the NIHR Oxford BRC. The views
expressed are those of the authors and not necessarily those of the NHS, the
NIHR or the Department of Health.

Conflict of Interest: The authors declare no conflict of interest.

\section*{Funding}
This work was supported by the National Institutes of Health (NIH) under Award Numbers
H6R00550\_CS00.01. 

\newpage
\bibliographystyle{plainnat}
\bibliography{references}

\begin{thebibliography}{26}
\providecommand{\natexlab}[1]{#1}
\providecommand{\url}[1]{\texttt{#1}}
\expandafter\ifx\csname urlstyle\endcsname\relax
  \providecommand{\doi}[1]{doi: #1}\else
  \providecommand{\doi}{doi: \begingroup \urlstyle{rm}\Url}\fi

\bibitem[Barndorff-Nielsen and Yeo(1969)]{barndorff1969negative}
Ole Barndorff-Nielsen and GF~Yeo.
\newblock Negative binomial processes.
\newblock \emph{Journal of Applied Probability}, 6\penalty0 (3):\penalty0
  633--647, 1969.

\bibitem[Benjamini and Hochberg(1995)]{benjamini1995controlling}
Yoav Benjamini and Yosef Hochberg.
\newblock Controlling the false discovery rate: a practical and powerful
  approach to multiple testing.
\newblock \emph{Journal of the Royal statistical society: series B
  (Methodological)}, 57\penalty0 (1):\penalty0 289--300, 1995.

\bibitem[Eickhoff et~al.(2012)Eickhoff, Bzdok, Laird, Kurth, and
  Fox]{eickhoff2012activation}
Simon~B Eickhoff, Danilo Bzdok, Angela~R Laird, Florian Kurth, and Peter~T Fox.
\newblock Activation likelihood estimation meta-analysis revisited.
\newblock \emph{Neuroimage}, 59\penalty0 (3):\penalty0 2349--2361, 2012.

\bibitem[Eisenberg et~al.(1966)Eisenberg, Geoghagen, and
  Walsh]{eisenberg1966general}
Herbert~B Eisenberg, Randolph~RM Geoghagen, and John~E Walsh.
\newblock A general use of the poisson approximation for binomial events, with
  application to bacterial endocarditis data.
\newblock \emph{Biometrics}, pages 74--82, 1966.

\bibitem[Gallagher et~al.(2019)Gallagher, Bulteau, Cohen, and
  Michaud]{gallagher2019neurocognitive}
Anne Gallagher, Christine Bulteau, David Cohen, and Jacques~L Michaud.
\newblock \emph{Neurocognitive Development: Normative Development}.
\newblock Elsevier, 2019.

\bibitem[Geoffroy and Weerakkody(2001)]{geoffroy2001poisson}
Pedro Geoffroy and Govinda Weerakkody.
\newblock A poisson-gamma model for two-stage cluster sampling data.
\newblock \emph{Journal of Statistical Computation and Simulation}, 68\penalty0
  (2):\penalty0 161--172, 2001.

\bibitem[Goudie et~al.(2019)Goudie, Presanis, Lunn, De~Angelis, and
  Wernisch]{goudie2019joining}
Robert~JB Goudie, Anne~M Presanis, David Lunn, Daniela De~Angelis, and Lorenz
  Wernisch.
\newblock Joining and splitting models with markov melding.
\newblock \emph{Bayesian analysis}, 14\penalty0 (1):\penalty0 81, 2019.

\bibitem[Hill-Bowen et~al.(2021)Hill-Bowen, Riedel, Poudel, Salo, Flannery,
  Camilleri, Eickhoff, Laird, and Sutherland]{hill2021cue}
Lauren~D Hill-Bowen, Michael~C Riedel, Ranjita Poudel, Taylor Salo, Jessica~S
  Flannery, Julia~A Camilleri, Simon~B Eickhoff, Angela~R Laird, and Matthew~T
  Sutherland.
\newblock The cue-reactivity paradigm: An ensemble of networks driving
  attention and cognition when viewing drug and natural reward-related stimuli.
\newblock \emph{Neuroscience \& Biobehavioral Reviews}, 130:\penalty0 201--213,
  2021.

\bibitem[Kang et~al.(2011)Kang, Johnson, Nichols, and Wager]{kang2011meta}
Jian Kang, Timothy~D Johnson, Thomas~E Nichols, and Tor~D Wager.
\newblock Meta analysis of functional neuroimaging data via bayesian spatial
  point processes.
\newblock \emph{Journal of the American Statistical Association}, 106\penalty0
  (493):\penalty0 124--134, 2011.

\bibitem[Kang et~al.(2014)Kang, Nichols, Wager, and Johnson]{kang2014bayesian}
Jian Kang, Thomas~E Nichols, Tor~D Wager, and Timothy~D Johnson.
\newblock A bayesian hierarchical spatial point process model for multi-type
  neuroimaging meta-analysis.
\newblock \emph{The annals of applied statistics}, 8\penalty0 (3):\penalty0
  1800, 2014.

\bibitem[Laird et~al.(2005)Laird, Lancaster, and Fox]{laird2005brainmap}
Angela~R Laird, Jack~J Lancaster, and Peter~T Fox.
\newblock Brainmap.
\newblock \emph{Neuroinformatics}, 3\penalty0 (1):\penalty0 65--77, 2005.

\bibitem[Lawless(1987)]{lawless1987negative}
Jerald~F Lawless.
\newblock Negative binomial and mixed poisson regression.
\newblock \emph{The Canadian Journal of Statistics/La Revue Canadienne de
  Statistique}, pages 209--225, 1987.

\bibitem[Longford(1987)]{longford1987fast}
Nicholas~T Longford.
\newblock A fast scoring algorithm for maximum likelihood estimation in
  unbalanced mixed models with nested random effects.
\newblock \emph{Biometrika}, 74\penalty0 (4):\penalty0 817--827, 1987.

\bibitem[Montagna et~al.(2018)Montagna, Wager, Barrett, Johnson, and
  Nichols]{montagna2018spatial}
Silvia Montagna, Tor Wager, Lisa~Feldman Barrett, Timothy~D Johnson, and
  Thomas~E Nichols.
\newblock Spatial bayesian latent factor regression modeling of
  coordinate-based meta-analysis data.
\newblock \emph{Biometrics}, 74\penalty0 (1):\penalty0 342--353, 2018.

\bibitem[Radua et~al.(2012)Radua, Mataix-Cols, Phillips, El-Hage, Kronhaus,
  Cardoner, and Surguladze]{radua2012new}
J~Radua, D~Mataix-Cols, Mary~L Phillips, W~El-Hage, DM~Kronhaus, N~Cardoner,
  and S~Surguladze.
\newblock A new meta-analytic method for neuroimaging studies that combines
  reported peak coordinates and statistical parametric maps.
\newblock \emph{European psychiatry}, 27\penalty0 (8):\penalty0 605--611, 2012.

\bibitem[Salimi-Khorshidi et~al.(2009)Salimi-Khorshidi, Smith, Keltner, Wager,
  and Nichols]{salimi2009meta}
Gholamreza Salimi-Khorshidi, Stephen~M Smith, John~R Keltner, Tor~D Wager, and
  Thomas~E Nichols.
\newblock Meta-analysis of neuroimaging data: a comparison of image-based and
  coordinate-based pooling of studies.
\newblock \emph{Neuroimage}, 45\penalty0 (3):\penalty0 810--823, 2009.

\bibitem[Samartsidis et~al.(2017{\natexlab{a}})Samartsidis, Montagna, Laird,
  Fox, Johnson, and Nichols]{samartsidis2017estimating}
Pantelis Samartsidis, Silvia Montagna, Angela~R Laird, Peter~T Fox, Timothy~D
  Johnson, and Thomas~E Nichols.
\newblock Estimating the number of missing experiments in a neuroimaging
  meta-analysis.
\newblock \emph{BioRxiv}, page 225425, 2017{\natexlab{a}}.

\bibitem[Samartsidis et~al.(2017{\natexlab{b}})Samartsidis, Montagna, Nichols,
  and Johnson]{samartsidis2017coordinate}
Pantelis Samartsidis, Silvia Montagna, Thomas~E Nichols, and Timothy~D Johnson.
\newblock The coordinate-based meta-analysis of neuroimaging data.
\newblock \emph{Statistical science: a review journal of the Institute of
  Mathematical Statistics}, 32\penalty0 (4):\penalty0 580, 2017{\natexlab{b}}.

\bibitem[Samartsidis et~al.(2019)Samartsidis, Eickhoff, Eickhoff, Wager,
  Barrett, Atzil, Johnson, and Nichols]{samartsidis2019bayesian}
Pantelis Samartsidis, Claudia~R Eickhoff, Simon~B Eickhoff, Tor~D Wager,
  Lisa~Feldman Barrett, Shir Atzil, Timothy~D Johnson, and Thomas~E Nichols.
\newblock Bayesian log-gaussian cox process regression: with applications to
  meta-analysis of neuroimaging working memory studies.
\newblock \emph{Journal of the Royal Statistical Society. Series C, Applied
  statistics}, 68\penalty0 (1):\penalty0 217, 2019.

\bibitem[Shanno(1970)]{shanno1970conditioning}
David~F Shanno.
\newblock Conditioning of quasi-newton methods for function minimization.
\newblock \emph{Mathematics of computation}, 24\penalty0 (111):\penalty0
  647--656, 1970.

\bibitem[Turkeltaub et~al.(2002)Turkeltaub, Eden, Jones, and
  Zeffiro]{turkeltaub2002meta}
Peter~E Turkeltaub, Guinevere~F Eden, Karen~M Jones, and Thomas~A Zeffiro.
\newblock Meta-analysis of the functional neuroanatomy of single-word reading:
  method and validation.
\newblock \emph{Neuroimage}, 16\penalty0 (3):\penalty0 765--780, 2002.

\bibitem[Ver~Hoef and Boveng(2007)]{ver2007quasi}
Jay~M Ver~Hoef and Peter~L Boveng.
\newblock Quasi-poisson vs. negative binomial regression: how should we model
  overdispersed count data?
\newblock \emph{Ecology}, 88\penalty0 (11):\penalty0 2766--2772, 2007.

\bibitem[Wager et~al.(2007)Wager, Lindquist, and Kaplan]{wager2007meta}
Tor~D Wager, Martin Lindquist, and Lauren Kaplan.
\newblock Meta-analysis of functional neuroimaging data: current and future
  directions.
\newblock \emph{Social cognitive and affective neuroscience}, 2\penalty0
  (2):\penalty0 150--158, 2007.

\bibitem[Westfall and Young(1993)]{westfall1993resampling}
Peter~H Westfall and S~Stanley Young.
\newblock \emph{Resampling-based multiple testing: Examples and methods for
  p-value adjustment}, volume 279.
\newblock John Wiley \& Sons, 1993.

\bibitem[Yarkoni et~al.(2011)Yarkoni, Poldrack, Nichols, Van~Essen, and
  Wager]{yarkoni2011large}
Tal Yarkoni, Russell~A Poldrack, Thomas~E Nichols, David~C Van~Essen, and Tor~D
  Wager.
\newblock Large-scale automated synthesis of human functional neuroimaging
  data.
\newblock \emph{Nature methods}, 8\penalty0 (8):\penalty0 665--670, 2011.

\bibitem[Yue et~al.(2012)Yue, Lindquist, and Loh]{yue2012meta}
Yu~Ryan Yue, Martin~A Lindquist, and Ji~Meng Loh.
\newblock Meta-analysis of functional neuroimaging data using bayesian
  nonparametric binary regression.
\newblock \emph{The Annals of Applied Statistics}, pages 697--718, 2012.

\end{thebibliography}

\newpage
\appendix
\section*{Supplementary Material}

\renewcommand{\thesection}{S\arabic{section}}
\renewcommand{\thetable}{S\arabic{table}}
\renewcommand{\thefigure}{S\arabic{figure}}

\section{Detailed derivation of stochastic models}
\subsection{Poisson model}
\label{appendix: Poisson model}
We assert that the sum of two independent Poisson random variables is also Poisson. Let $X \sim \mathrm{Poi}(\lambda_1)$ and $Y \sim \mathrm{Poi}(\lambda_2)$ be two independent random variables, and $Z=X+Y$, then,
\begin{equation}
\begin{split}
    P(Z=n) = P(X+Y=N) &= \sum \limits_{k=-\infty}^\infty P(X=k) P(Y=n-K) \\
    &= \sum \limits_{k=0}^n P(X=k) P(Y=n-k) \\
    &= \sum \limits_{k=0}^n e^{-\lambda_1} \frac{\lambda_1^k}{k!} e^{-\lambda_2} \frac{\lambda_2^{n-k}}{(n-k)!} \\
    &= e^{-(\lambda_1 + \lambda_2)} \sum \limits_{k=0}^n \frac{\lambda_1^k \lambda_2^{n-k}}{k!(n-k)!} \\
    &= \frac{e^{-(\lambda_1+\lambda_2)}}{n!} \sum \limits_{k=0}^n \frac{n!}{k!(n-k)!} \lambda_1^k \lambda_2^{n-k} \\
    &= \frac{e^{-(\lambda_1+\lambda_2)}}{n!} (\lambda_1+\lambda_2)^n
\end{split}
\end{equation}
Therefore, $Z=X+Y \sim \mathrm{Poi}(\lambda_1 + \lambda_2)$ is also a Poisson variable. The conclusion can be further extended: the sum of multiple Poisson random variables ($\mathrm{Poi}(\lambda_i), i=1, \cdots, n$) also follows Poisson distribution, with the parameter $\lambda=\sum \limits_{i=1}^n \lambda_i$.

Hence, under the assumption of independence of counts across studies, we believe that the likelihood function is exactly same if we model the voxelwise total foci count over studies (with length-N) instead of voxelwise foci count for each study (with expanded length-(NM)). This reformulation can simplify the computation of the log-likelihood function and reduce the dimensionality of statistics (never larger than $M$ or $N$ in dimension).

\subsection{Negative Binomial (Poisson-Gamma) Model}
\label{appendix: Negative Binomial (Poisson-Gamma) model}
In this section, we describe the formulation of NB distribution in more detail. Based on the assumption of NB (Poisson-Gamma) model, there's a single parameter $\alpha$ indicates variance in excess of Poisson model, for voxel $j$ in study $i$, the voxelwise mean of intensity $\lambda_{ij}$ follows a Gamma distribution with mean $\mathbb{E}(\lambda_{ij}) = \mu_{ij}$ and variance $\mathrm{Var}(\lambda_{ij})=\alpha \mu_{ij}^2$
\begin{equation*}
    \lambda_{ij} \sim \mathrm{Gamma}(\alpha^{-1}, \frac{\alpha^{-1}}{\mu_{ij}})
    \Rightarrow \mathbb{E}(\lambda_{ij}) = \mu_{ij}, 
    \mathrm{Var}(\lambda_{ij}) = \alpha \mu_{ij}^2
\end{equation*}
And $Y_{ij}\vert \lambda_{ij}$ follows a Poisson distribution with conditional mean $\mathbb{E}(Y_{ij}\vert \lambda_{ij}) = \lambda_{ij}$
\begin{equation*}
Y_{ij} \vert \lambda_{ij} = \mathrm{Poisson}(\lambda_{ij}) 
\Rightarrow P(Y_{ij} \vert \lambda_{ij} = k) = \frac{\lambda_{ij}^{k}e^{-\lambda_{ij}}}{k!}
\end{equation*}
which gives rise to marginal probability of $Y_{ij}$
\begin{equation*}
    \begin{split}
        P(Y_{ij} = y_{ij}) &= \int \limits_{\lambda_{ij}} P(Y_{ij} \vert \lambda_{ij}) P(\lambda_{ij}) d \lambda_{ij} 
        = \int \limits_{\lambda_{ij}=0}^{\infty} \frac{\lambda_{ij}^{y_{ij}} e^{-\lambda_{ij}}}{y_{ij} !} \frac{(\frac{1}{\alpha \mu_{ij}})^\frac{1}{\alpha}}{\Gamma(\frac{1}{\alpha})} \lambda_{ij}^{\frac{1}{\alpha}-1} e^{-\frac{\lambda_{ij}}{\alpha \mu_{ij}}} d \lambda_{ij} \\
        &= \frac{\frac{1}{\alpha \mu_{ij}}^\frac{1}{\alpha}}{y_{ij}! \Gamma(\frac{1}{\alpha})} \int \limits_{\lambda_{ij}=0}^{\infty} \lambda_{ij}^{y_{ij}} e^{-\lambda_{ij}} \lambda_{ij}^{\frac{1}{\alpha}-1} e^{-\frac{\lambda_{ij}}{\alpha \mu_{ij}}} d \lambda_{ij} 
        = \frac{\frac{1}{\alpha \mu_{ij}}^\frac{1}{\alpha}}{y_{ij}! \Gamma(\frac{1}{\alpha})} \frac{\Gamma(y_{ij}+\frac{1}{\alpha})}{(\frac{1}{\alpha \mu_{ij}}+1)^{y_{ij}+\frac{1}{\alpha}}} \\
        &= \frac{\Gamma(y_{ij}+\frac{1}{\alpha})}{\Gamma(y_{ij}+1) \Gamma(\frac{1}{\alpha})} (\frac{\frac{1}{\alpha \mu_{ij}}}{\frac{1}{\alpha \mu_{ij}}+1})^{\frac{1}{\alpha}} (\frac{1}{\frac{1}{\alpha \mu_{ij}}+1})^{y_{ij}}
        = \frac{\Gamma(y_{ij}+\alpha^{-1})}{\Gamma(y_{ij}+1) \Gamma(\alpha^{-1})} (\frac{1}{1+\alpha \mu_{ij}})^{\alpha^{-1}} (\frac{\alpha \mu_{ij}}{1+\alpha \mu_{ij}})^{y_{ij}}
    \end{split}
\end{equation*}
which satisfies the mathematical form of probability density function of NB model, $Y_{ij} \sim NB(\alpha^{-1}, \frac{\mu_{ij}}{\alpha^{-1}+\mu_{ij}})$, with mean $\mathbb{E}[Y_{ij}] = \mu_{ij}$ and variance $\mathbb{V}(Y_{ij}) = \mu_{ij} + \alpha \mu_{ij}^2$.

\subsection{Moment Matching Approach}
\label{appendix: Moment Matching Approach}
For the purpose of approximating the sum of multiple independent NB random variables, we approximate a sum of NB variates with a NB distribution by moment matching (mean and variance). 
Suppose the voxelwise count in each individual study is $Y_{ij} \sim \mathrm{NB}(\alpha^{-1},\frac{\mu_{ij}}{\mu_{ij}+\alpha^{-1}})$, and $\alpha$ is global dispersion parameter. Using the independence of studies at voxel $j$,
\begin{equation*}
\begin{cases}
    \mathbb{E}(Y_{\centerdot,j}) = \sum \limits_{i=1}^M \mathbb{E}(Y_{ij}) = \sum \limits_{i=1}^M \mu_{ij} \\
    \mathbb{V}(\mathbb{E}(Y_{\centerdot,j})) = \sum \limits_{i=1}^M \mathrm{Var}(Y_{ij}) = \sum \limits_{i=1}^M \mu_{ij} + \sum \limits_{i=1}^M \alpha \mu_{ij}^2
\end{cases}
\end{equation*}

To ensure that the proposed NB distribution ($Y_{\centerdot,j} \sim NB(r', p')$) matches the mixture of NB distributions, with regard to both mean and variance, we need
\begin{equation*}
\label{eq: NB moment matching}
    \begin{cases}
      \mathbb{E}(Y_{\centerdot,j}) = \sum \limits_{i=1}^M \mu_{ij} \\
      \mathrm{Var}(Y_{\centerdot,j}) = \sum \limits_{i=1}^M \mu_{ij} + \sum \limits_{i=1}^M \alpha \mu_{ij}^2
    \end{cases}
\Rightarrow \begin{cases}
      p' = \frac{\sum \limits_{i=1}^M \mu_{ij}^2}{\alpha^{-1} \sum \limits_{i=1}^M \mu_{ij} + \sum \limits_{i=1}^M \mu_{ij}^2} \\
      r' = \frac{(\sum \limits_{i=1}^M \mu_{ij})^2}{\alpha \sum \limits_{i=1}^M \mu_{ij}^2}
    \end{cases}
\end{equation*}
Therefore, approximated NB distribution of sum of foci count at voxel $j$ is, $Y_{\centerdot,j} \sim  \mathrm{NB} \left(\frac{(\sum \limits_{i=1}^M \mu_{ij})^2}{\alpha \sum \limits_{i=1}^M \mu_{ij}^2}, \frac{\sum \limits_{i=1}^M \mu_{ij}^2}{v \sum \limits_{i=1}^M \mu_{ij} + \sum \limits_{i=1}^M \mu_{ij}^2} \right)$, with excess variance in NB approximation $\alpha'$,
\begin{equation*}
\frac{1}{\alpha'} = \frac{(\sum \limits_{i=1}^M \mu_{ij})^2}{\alpha \sum \limits_{i=1}^M \mu_{ij}^2}
\Rightarrow \alpha' = \frac{\sum \limits_{i=1}^M \mu_{ij}^2}{(\sum \limits_{i=1}^M \mu_{ij})^2}  \alpha
\end{equation*}. 

\subsection{Two-stage hierarchy Poisson-Gamma model}
\label{appendix: Two-stage hierarchy Poisson-Gamma model}
In this section, we propose a two-stage hierarchy Poisson-Gamma model, which regards random (Gamma) effect as a latent characteristic of each study, instead of independent voxelwise effects. The name ``two-stage hierarchy Poisson-Gamma model" comes from the modelling procedure: consider the clustered count data $Y_{ij}$, $i=1, \cdots, M$ (number of studies), $j=1,\cdots, N$ (number of voxels). Draw $\lambda_i$ from a Gamma distribution with mean $1$ and variance $\alpha$. 
\begin{equation*}
        \lambda_i \sim \mathrm{Gamma}(\alpha^{-1}, \alpha^{-1})
        \Rightarrow \mathbb{E}(\lambda_i) = 1,
        \mathbb{V}(\lambda_i) = \alpha,
\end{equation*}
\begin{equation*}
    f(\lambda_i) = \frac{\frac{1}{\alpha}^{\alpha^{-1}}}{\Gamma(\alpha^{-1})} \lambda_i^{\alpha-1} e^{-\alpha^{-1} \lambda_i}
\end{equation*}
For each study $i$, for each voxel $j$, $Y_{ij} \vert \lambda_i$ are drawn from a Poisson distribution with mean $\lambda_i \mu_{ij}$, where $\mu_{ij}$ is the spatial mean parameterized by some $\beta$ (B-spline basis coefficients). 
\begin{equation*}
        Y_{ij} \vert \lambda_i \sim \mathrm{Poisson}(\lambda_i \mu_{ij})
        \Rightarrow P(Y_{ij} \vert \lambda_i=k) = \frac{(\lambda_i \mu_{ij})^k e^{-\lambda_i \mu_{ij}}}{k!}
\end{equation*}
Therefore, marginal probability of foci count $Y_{ij}$ is,
\begin{equation*}
    \begin{split}
         P(Y_{ij}=y_{ij}) &= \int \limits_{\lambda_i} P(Y_{ij}\vert \lambda_i) P(\mu_i) d \lambda_i
        = \int \limits_{\lambda_i=0}^{\infty} \frac{(\mu_{ij} \lambda_i)^{y_{ij}} e^{-\mu_{ij} \lambda_i}}{y_{ij}!}  \frac{\frac{1}{\alpha}^{\alpha^{-1}}}{\Gamma(\alpha^{-1})} \lambda_i^{\alpha^{-1}-1} e^{-\alpha^{-1} \lambda_i} d \lambda_i \\
        &= \frac{\mu_{ij}^{y_{ij}} \frac{1}{\alpha}^{\alpha^{-1}}}{\Gamma(\alpha^{-1}) y_{ij}!} \int \limits_{\lambda_i=0}^{\infty} \lambda_i^{y_{ij}+\alpha^{-1}-1} e^{-\lambda_i(\mu_{ij}+\alpha^{-1})} d \lambda_i 
        = \frac{\mu_{ij}^{y_{ij}} \frac{1}{\alpha}^{\alpha^{-1}}}{\Gamma(\alpha^{-1}) y_{ij}!} \frac{\Gamma(y_{ij}+\alpha^{-1})}{(\mu_{ij}+\alpha^{-1})^{y_{ij}+\alpha^{-1}}} \\
        &= \frac{\Gamma(y_{ij}+\alpha^{-1})}{\Gamma(y_{ij}+1) \Gamma(\alpha^{-1})} (\frac{\mu_{ij}}{\mu_{ij}+\alpha^{-1}})^{y_{ij}} (\frac{\alpha^{-1}}{\mu_{ij}+\alpha^{-1}})^{\alpha^{-1}}
    \end{split}
\end{equation*}
Therefore, marginal distribution of foci count follows a NB distribution, $Y_{ij} \sim \mathrm{NB}(\alpha^{-1}, \frac{\mu_{ij}}{\mu_{ij}+\alpha^{-1}})$, with mean $\mathbb{E}(Y_{ij}) = \mu_{ij}$ and variance $\mathrm{Var}(Y_{ij}) = \mu_{ij} + \alpha \mu_{ij}^2$.

\subsection{Covariance structure in Clustered NB model}
\label{appendix: Covariance structure in Clustered NB model}
The two-stage hierarchical clustered NB model also introduces a covariance structure between foci within a study, specifically, covariance of number of count in voxel $j$ and voxel $j'$ ($Y_{ij}$ and $Y_{ij'}$) in study $i$, modelled by clustered NB model is, 
\begin{equation*}
    \begin{split}
        \mathbb{E}[Y_{ij} Y_{ij'}] &= \mathbb{E}_{\lambda_i} \left[\mathbb{E}[Y_{ij}Y_{ij'}\vert \lambda_i]\right] = \mathbb{E}_{\lambda_i} \left[\mathbb{E}[Y_{ij} \vert \lambda_i]  \mathbb{E}[Y_{ij'} \vert \lambda_i]\right] 
        = \mathbb{E}_{\lambda_i} \left[(\lambda_i  \mu_{ij})(\lambda_i  \mu_{ij'}) \right] \\
        &= \mu_{ij}  \mu_{ij'} \int_{\lambda_i} \lambda_i^2 f(\lambda_i) d \lambda_i 
        = \mu_{ij}  \mu_{ij'} \int_{\lambda_i} \frac{\frac{1}{\alpha}^{\alpha^{-1}}}{\Gamma(\alpha^{-1})} \lambda_i^{\alpha^{-1}+1} e^{-\alpha^{-1} \lambda_i} d \lambda_i \\
        &= \mu_{ij}  \mu_{ij'}  \frac{\frac{1}{\alpha}^{\alpha^{-1}}}{\Gamma(\alpha^{-1})}   \alpha^{\alpha^{-1}+1}  \alpha \int_{\lambda_i} (\frac{1}{\alpha}\lambda_i)^{(\alpha^{-1}+1)} e^{\alpha^{-1} \lambda_i} d(\alpha^{-1} \lambda_i) \\
        &= \mu_{ij}  \mu_{ij'}  \alpha^2  \frac{1}{\Gamma(\alpha^{-1})} \Gamma(\alpha^{-1}+2) 
        = \mu_{ij}  \mu_{ij'}  \alpha^2  \alpha^{-1} (\alpha^{-1} + 1) \\
        &= (1+\alpha) \mu_{ij}  \mu_{ij'}
    \end{split}
\end{equation*}

For different voxels $j$ and $j'$ within a same study $i$,
\begin{equation*}
    \mathrm{Cov}(Y_{ij}Y_{ij'}) = \mathbb{E}[Y_{ij} Y_{ij'}] - \mathbb{E}[Y_{ij}]\mathbb{E}[Y_{ij'}]
    = (1+\alpha) \mu_{ij} \mu_{ij'} - \mu_{ij}  \mu_{ij'}
    = \alpha \mu_{ij}\mu_{ij'} 
\end{equation*}
While for different studies $i$ and $i'$,
\begin{equation*}
    \mathrm{Cov}(Y_{ij}Y_{i'j'}) = 0 
\end{equation*}

Now, we will look at the total log-likelihood function of clustered NB model, using dependence between studies. Let $Y_{i,\centerdot} = \sum \limits_{j=1}^N Y_{ij}$ be the sum of foci within study $i$ regardless of location. The joint probability for the number of count $Y_{ij} \, (\forall j=1, \cdots, N)$ from the $i^{th}$ study is,
\begin{equation*}
    \begin{split}
    & f(Y_{i1}, Y_{i2}, \cdots, Y_{iN}) = \int_{\lambda_i} f(Y_{i1}, Y_{i2}, \cdots, Y_{iN} \vert \lambda_i)  f(\lambda_i) d \lambda_i 
    = \int_{\lambda_i} \prod \limits_{j=1}^N f(Y_{ij} \vert \lambda_i)  f(\lambda_i) d \lambda_i\\
    &= \int_{\lambda_i} \prod \limits_{j=1}^N \frac{(\mu_{ij} \lambda_i)^{Y_{ij}}e^{-\mu_{ij} \lambda_i}}{Y_{ij}!}  \frac{\frac{1}{\alpha}^{\alpha^{-1}}}{\Gamma(\alpha^{-1})} \lambda_i^{\alpha^{-1}-1}e^{- \alpha^{-1} \lambda_i} d \lambda_i\\
    &= \frac{\frac{1}{\alpha}^{\alpha^{-1}} \prod \limits_{j=1}^N \mu_{ij}^{Y_{ij}}}{\Gamma(\alpha^{-1}) \prod \limits_{j=1}^N Y_{ij}!} \int_{\lambda_i} \exp\{-\lambda_i (\alpha^{-1}+\sum \limits_{j=1}^N \mu_{ij})\}  \lambda_i^{\sum \limits_{j=1}^N Y_{ij}+\alpha^{-1}-1} d \lambda_i \\
    &= \frac{\frac{1}{\alpha}^{\alpha^{-1}} \prod \limits_{j=1}^N \mu_{ij}^{Y_{ij}}}{\Gamma(\alpha^{-1}) \prod \limits_{j=1}^N Y_{ij}!} \int_{\lambda_i} \exp\{-\lambda_i (\alpha^{-1}+\mu_{i,\centerdot})\}  \lambda_i^{Y_{i,\centerdot}+\alpha^{-1}-1} d \lambda_i \\
    &= \frac{\frac{1}{\alpha}^{\alpha^{-1}} \prod \limits_{j=1}^N \mu_{ij}^{Y_{ij}}}{\Gamma(\alpha^{-1}) \prod \limits_{j=1}^N Y_{ij}!} \frac{1}{(\alpha^{-1}+\mu_{i,\centerdot})^{Y_{i,\centerdot}+\alpha^{-1}-1}}  \frac{1}{\alpha^{-1}+\mu_{i,\centerdot}} \cdot \\
    & \; \; \; \int_{\lambda_i} \exp\{-\lambda_i (\alpha^{-1}+\mu_{i,\centerdot})\}  [\lambda_i(\alpha^{-1}+\mu_{i,\centerdot})]^{Y_{i,\centerdot}+\alpha^{-1}-1} d[\lambda_i(\alpha^{-1}+\mu_{i,\centerdot})] \\
    &= \frac{\prod \limits_{j=1}^N \mu_{ij}^{Y_{ij}}}{\Gamma(\alpha^{-1})\prod \limits_{j=1}^N Y_{ij}!}  \frac{\frac{1}{\alpha}^{\alpha^{-1}}}{(\alpha^{-1}+\mu_{i,\centerdot})^{Y_{i,\centerdot}+\alpha^{-1}}} \Gamma(Y_{i,\centerdot}+\alpha^{-1}) \\
    &= \frac{\Gamma(Y_{i,\centerdot}+\alpha^{-1}) \frac{1}{\alpha}^{\alpha^{-1}}}{\Gamma(\alpha^{-1}) \prod \limits_{j=1}^N Y_{ij}!}  (\alpha^{-1}+\mu_{i,\centerdot})^{-(Y_{i,\centerdot}+\alpha^{-1})}  \exp{[\sum \limits_{j=1}^N Y_{ij} \log(\mu_{ij})]}
    \end{split}
\end{equation*}
It gives rise to log-likelihood function for study $i$, 
\begin{equation*}
\begin{split}
    \log f(Y_{i1}, \cdots, Y_{iN}) &= \alpha^{-1} \log(\alpha^{-1}) + \log \Gamma(Y_{i,\centerdot}+\alpha^{-1}) - \log\Gamma(\alpha^{-1}) - \sum \limits_{j=1}^N \log Y_{ij}! \\
    &- (Y_{i,\centerdot}+\alpha^{-1})\log(\alpha^{-1}+\mu_{i,\centerdot}) + \sum \limits_{j=1}^N Y_{ij} \log(\mu_{ij})
\end{split}
\end{equation*}

Therefore, using the independence between study $i$ and $i'$ ($i \neq i'$), the log-likelihood of $Y_{ij}$($\forall i=1,\cdots, M, j=1,\cdots,N$) across all studies and voxels is,
\begin{equation*}
    \begin{split}
        l(\beta, \alpha) &= \sum \limits_{i=1}^M \log[f(Y_{i1}, Y_{i2}, \cdots, Y_{iN})] \\
        &= M\alpha^{-1}\log(\alpha^{-1})-M\log\Gamma(\alpha^{-1}) + \sum \limits_{i=1}^M \log\Gamma(Y_{i,\centerdot}+\alpha^{-1}) - \sum \limits_{i=1}^M \sum\limits_{j=1}^N \log Y_{ij}! \\
        &- \sum \limits_{i=1}^M (Y_{i,\centerdot}+\alpha^{-1})\log(\alpha^{-1}+\mu_{i,\centerdot}) + \sum \limits_{i=1}^M \sum \limits_{j=1}^N Y_{ij} \log(\mu_{ij})
    \end{split}
\end{equation*}

Using the assumption that $Y_{ij}=0 \text{ or }1$, so that $\log(Y_{ij}!)=0$ and $\mu_{ij}=\mu_{j}^X  \mu_i^Z$,
\begin{equation*}
\label{equ: basis cluster NB log-likelihood}
    \begin{split}
        l(\beta, \alpha) &= M \alpha^{-1} \log(\alpha^{-1})-M\log\Gamma(\alpha^{-1}) + \sum \limits_{i=1}^M \log\Gamma(Y_{i,\centerdot}+\alpha^{-1}) - \sum \limits_{i=1}^M (Y_{i,\centerdot}+\alpha^{-1})\log(\alpha^{-1}+\mu_{i,\centerdot}) + \sum \limits_{i=1}^M \sum \limits_{j=1}^N Y_{ij} \log(\mu_{ij}) \\
        &= M \alpha^{-1} \log(\alpha^{-1})-M\log\Gamma(\alpha^{-1}) + \sum \limits_{i=1}^M \log\Gamma(Y_{i,\centerdot}+\alpha^{-1}) -  \sum \limits_{i=1}^M (Y_{i,\centerdot}+\alpha^{-1})\log(\alpha^{-1}+\mu_{i,\centerdot}) \\
        &+ \left(\sum \limits_{j=1}^N \sum \limits_{i=1}^M Y_{ij}\right)  \left(\sum_{k=1}^P X_{jk}\beta_{k}\right) \\
        &= M\alpha^{-1} \log(\alpha^{-1})-M\log\Gamma(\alpha^{-1}) + \sum \limits_{i=1}^M \log\Gamma(Y_{i,\centerdot}+\alpha^{-1}) - \sum \limits_{i=1}^M (Y_{i,\centerdot}+\alpha^{-1})\log(\alpha^{-1}+\mu_{i,\centerdot}) \\
        &+ \left[\sum \limits_{j=1}^N Y_{\centerdot,j} \sum_{k=1}^P X_{jk}\beta_{k}\right] \\
        &=  M\alpha^{-1} \log(\alpha^{-1})-M\log\Gamma(\alpha^{-1}) + \sum \limits_{i=1}^M \log\Gamma(Y_{i,\centerdot}+\alpha^{-1}) - \sum \limits_{i=1}^M (Y_{i,\centerdot}+\alpha^{-1})\log(\alpha^{-1}+\mu_{i,\centerdot}) 
        + Y_{\centerdot,}^{\top} \log(\mu^X)
    \end{split}
\end{equation*}

\section{Deterministic model}
\label{appendix: Model factorisation}
\subsection{Model factorisation: Poisson model}
\label{appendix: Model factorisation: Poisson model}
We consider model factorisation to replace the full $(MN)-vector$ of foci count by sufficient statistics (dimension not larger than $M$ or $N$). Following the total log-likelihood function in \Cref{eq: basic Poisson likelihood}, 
\begin{equation}
\begin{split}
    l &= \sum \limits_{i=1}^{M} \sum \limits_{j=1}^{N} [Y_{ij} \log(\mu_{ij}) - \mu_{ij} - \log(y_{ij}!)] 
    = \sum \limits_{i=1}^M \sum \limits_{j=1}^N Y_{ij} \log{\mu_{ij}} - \sum \limits_{i=1}^M \sum \limits_{j=1}^N \mu_{ij} - 0 \\
    &= \left( \sum \limits_{i=1}^M \sum \limits_{j=1}^N Y_{ij} \right)  \left(\sum \limits_{k=1}^P X_{jk} \beta_k + \sum \limits_{s=1}^R Z_{is} \gamma_s\right) - \sum \limits_{i=1}^M \sum \limits_{j=1}^N  \mu^X_{j} \mu^Z_i \\
    &= \left[\sum \limits_{j=1}^N Y_{\centerdot,j} \sum \limits_{k=1}^P X_{jk} \beta_{gk}\right] + \sum \limits_{i=1}^M Y_{i,\centerdot} \sum \limits_{s=1}^R Z_{is} \gamma_s -  \left[\sum \limits_{j=1}^N \mu^X_{j}\right] \left[\sum \limits_{i=1}^M \mu^Z_i\right] \\
    &=\left[ \sum \limits_{j=1}^N Y_{\centerdot,j} \log\mu^X_{j}\right] + \sum \limits_{i=1}^M Y_{i,\centerdot} \log\mu^Z_i - \left[ \boldsymbol{1}^{\top} \mu^X \right] \left[\boldsymbol{1}^{\top} \mu^Z \right] \\
    &= Y_{\centerdot,}^{\top} \log(\mu^X) + Y_{,\centerdot}^{\top} \log(\mu^Z) - \left[ \boldsymbol{1}^{\top} \mu^X \right] \left[\boldsymbol{1}^{\top} \mu^Z \right]
\end{split}
\end{equation}

\subsection{Model factorisation: NB model}
\label{Model factorisation: NB model}
Following the log-likelihood function in \Cref{equation: pdf of NB model}, 
\begin{equation}
\begin{split}
    l(\beta, \alpha) &= \sum \limits_{i=1}^{M} \sum \limits_{j=1}^{N} \left[\log \Gamma(Y_{ij} + \alpha^{-1}) - \log \Gamma(Y_{ij}+1) - \log \Gamma(\alpha^{-1}) + Y_{ij} \log(\alpha \mu_{ij}) - (Y_{ij}+\alpha^{-1})\log(1+\alpha \mu_{ij})\right]\\
    &= \sum \limits_{i=1}^{M} \sum \limits_{j=1}^{N} \left[ \{\sum \limits_{k=0}^{Y_{ij}-1} \log(k+\alpha^{-1})\} - \log\Gamma(Y_{ij}+1) + Y_{ij} \log(\alpha \mu_{ij}) - (Y_{ij}+\alpha^{-1})\log(1+\alpha \mu_{ij})\right] \\
    &= \sum \limits_{i=1}^{M} \sum \limits_{j=1}^{N} \left[ \{\sum \limits_{k=0}^{Y_{ij}-1} \log(k+\alpha^{-1})\} - \log\Gamma(Y_{ij}+1) + Y_{ij} \log(\alpha) + Y_{ij}\log(\mu_{ij}) - (Y_{ij}+\alpha^{-1}) \log(1+\alpha\mu_{ij}) \right] \\
    &=\left(\sum \limits_{i=1}^M \sum \limits_{j=1}^N Y_{ij}  \log(\alpha^{-1}) - \sum \limits_{i=1}^M \sum \limits_{j=1}^N \log(1) \right) + \left(\sum \limits_{i=1}^M \sum \limits_{j=1}^N Y_{ij} \right)  \log(\alpha)  \\
    & + \sum \limits_{i=1}^M \sum \limits_{j=1}^N Y_{ij} \left(\sum \limits_{k=1}^P X_{jk}\beta_{k} + \sum \limits_{s=1}^R Z_{is} \gamma_s\right) - \sum \limits_{i=1}^M \sum \limits_{j=1}^N (Y_{ij}+\alpha^{-1}) \log(1+\alpha \mu_{ij})
    \end{split}
    \end{equation}
Here, the last term $\sum \limits_{j=1}^N (Y_{ij}+\alpha^{-1}) \log(1+\alpha \mu_{ij})$ is impractical to simplify, therefore, we consider moment matching method similar as \Cref{eq: NB moment matching}, $Y_{\centerdot,j} \sim \mathrm{NB}(r'_j, p'_j)$ where
\begin{equation}
\begin{split}
    r'_j &= \alpha^{-1}  \frac{(\sum \limits_{i=1}^M \mu_{ij})^2}{\sum \limits_{i=1}^M \mu_{ij}^2} = \alpha^{-1}  \frac{(\mu^X_j \sum \limits_{i=1}^M  \mu^Z_i)^2}{\sum \limits_{i=1}^M (\mu^X_j  \mu^Z_i)^2} = \alpha^{-1}  \frac{(\mu^X_j)^2  (\sum \limits_{i =1}^M \mu^Z_i)^2}{\sum \limits_{i=1}^M (\mu^X_j  \mu^Z_i)^2}\\
    p'_j &= \frac{\sum \limits_{i=1}^M \mu_{ij}^2}{\alpha^{-1} \sum \limits_{i=1}^M \mu_{ij} + \sum \limits_{i=1}^M \mu_{ij}^2} = \frac{\sum \limits_{i=1}^M (\mu^X_j \mu^Z_i)^2}{\alpha^{-1} \sum \limits_{i=1}^M (\mu^X_j  \mu^Z_i)+\sum \limits_{i=1}^M (\mu^X_j  \mu^Z_i)^2} = \frac{\sum \limits_{i=1}^M (\mu^X_j \mu^Z_i)^2}{\alpha^{-1}  \mu^X_j \sum \limits_{i=1}^M \mu^Z_i+\sum \limits_{i =1}^M (\mu^X_j \mu^Z_i)^2}
\end{split}
\end{equation}
And the parameter $\alpha'$ of extra variance in NB approximation is

\begin{equation}
    \frac{1}{\alpha'} = \frac{1}{\alpha}  \frac{(\sum \limits_{i=1}^M \mu_{ij})^2}{\sum \limits_{i=1}^M \mu_{ij}^2}
    \Rightarrow \alpha' = \frac{\sum \limits_{i=1}^M \mu_{ij}^2}{(\sum \limits_{i=1}^M \mu_{ij})^2}  \alpha = \frac{\sum \limits_{i=1}^M(\mu^X_j \mu^Z_i)^2}{(\mu^X_j)^2(\sum \limits_{i=1}^M \mu^Z_i)^2}  \alpha
\end{equation}

\subsection{Model factorisation: clustered NB model}
\label{Model factorisation: clustered NB model}
Following  the  total  log-likelihood  function  in  \Cref{equ: basis cluster NB log-likelihood},  we  incorporate  the  effect  of  study-level covariates into the Clustered NB model,
\begin{equation}
    \begin{split}
        l(\beta, \alpha) &= Mv\log(v)-M\log\Gamma(v) + \sum \limits_{i=1}^M \log\Gamma(Y_{i,\centerdot}+v) - \sum \limits_{i=1}^M (Y_{i,\centerdot}+v)\log(v+\mu_{i,\centerdot}) + \sum \limits_{i=1}^M \sum \limits_{j=1}^N Y_{ij} \log(\mu_{ij}) \\
        &= Mv\log(v)-M\log\Gamma(v) + \sum \limits_{g=1}^B \sum \limits_{i \in I_g} \log\Gamma(Y_{i,\centerdot}+v) - \sum \limits_{g=1}^B \sum \limits_{i \in I_g} (Y_{i,\centerdot}+v)\log(v+\mu_{i,\centerdot}) \\
        &+ \sum_{g=1}^B \left(\sum \limits_{j=1}^N \sum \limits_{i \in I_g} Y_{ij}\right)  \left(\sum_{k=1}^P X_{jk}\beta_{g(i)k} + \sum \limits_{s=1}^R Z_{is} \gamma_s \right) \\
        &= Mv\log(v)-M\log\Gamma(v) + \sum \limits_{g=1}^B \sum \limits_{i \in I_g} \log\Gamma(Y_{i,\centerdot}+v) - \sum \limits_{g=1}^B \sum \limits_{i \in I_g} (Y_{i,\centerdot}+v)\log(v+\mu_{i,\centerdot}) \\
        &+ \sum_{g=1}^B \left[\sum \limits_{j=1}^N Y_{gj} \sum_{k=1}^P X_{jk}\beta_{g(i)k}\right] + \sum \limits_{i=1}^M Y_{i,\centerdot} \sum \limits_{s=1}^R Z_{is} \gamma_s \\
        &= Mv\log(v)-M\log\Gamma(v) + \sum \limits_{g=1}^B \sum \limits_{i \in I_g} \log\Gamma(Y_{i,\centerdot}+v) - \sum \limits_{g=1}^B \sum \limits_{i \in I_g} (Y_{i,\centerdot}+v)\log(v+\mu_{i,\centerdot}) \\
        &+ \sum \limits_{g=1}^B Y_g^{\top} \log(\mu_g^X) + Y_{,\centerdot}^{\top} \log(\mu^Z)
    \end{split}
\end{equation}

\section{Statistical inference and generalised linear hypothesis testing}
\subsection{Contrasts on regression coefficient of study-level covariates}
\label{Appendix: Contrasts on regression coefficient of study-level covariates}
To investigate the effects of study-level covariates (e.g., sample size, year of publication) on activation intensity estimation, we conduct generalised linear hypothesis testing (GLH) on the regression coefficients $\gamma$. For every study-level covariate $\gamma_r, \forall r=1,\cdots, s$, 
\begin{itemize}
    \item $H_0: C_{\gamma}\gamma = C_{\gamma}[\gamma_1, \gamma_2, \cdots, \gamma_s]^T = \bold{0}_{m \times 1}^T$ where $C_{\gamma}$ is the contrast matrix of size $m \times s (m \leq s)$
    \item $H_1: C_{\gamma}\gamma = C_{\gamma}[\gamma_1, \gamma_2, \cdots, \gamma_s]^T \neq \bold{0}_{m \times 1}^T$
\end{itemize}
The covariance of regression coefficient $\gamma$, $\mathrm{Cov}_\gamma = \mathrm{Cov}([\gamma_1, \gamma_2, \cdots, \gamma_s]^T)$ is approximated from the inverse of Fisher Information matrix. According to asymptotic normality of maximum likelihood estimator, 
\begin{equation*}
\begin{split}
    \hat{\gamma} - \gamma &\xrightarrow{D} N \left(\bold{0}^T_{s \times1}, \mathrm{Cov}_\gamma \right) \\
    C_\gamma (\hat{\gamma} - \gamma) &\xrightarrow{D} N \left(\bold{0}^T_{s \times1}, C_\gamma \mathrm{Cov}_\gamma C_\gamma^T \right) \\
    C_\gamma \hat{\gamma} & \xrightarrow{D} N \left(C_{\gamma} \gamma, C_\gamma \mathrm{Cov}_\gamma C_\gamma^T \right)
\end{split}
\end{equation*}

Since a quadratic form of normal distribution has a Chi-square distribution,
\begin{equation*}
    (C_\gamma \hat{\gamma})^T (C_\gamma \mathrm{Cov}_\gamma C^T_\gamma)^{-1} (C_\gamma \hat{\gamma}) \xrightarrow{D} \chi^2_m
\end{equation*}
e.g., contrast matrix $C_\gamma = [1, 0]$ or $[0, 1]$ is for testing if the regression coefficient of the $1^{st}$ or $2^{nd}$ study-level covariate is zero.

\subsection{PP-plots of spatial homogeneity tests for each $20$ meta-analytic datasets}
\label{Appendix: validity of PP-plots}
\begin{figure}[h]
\centering
\includegraphics[width=15cm]{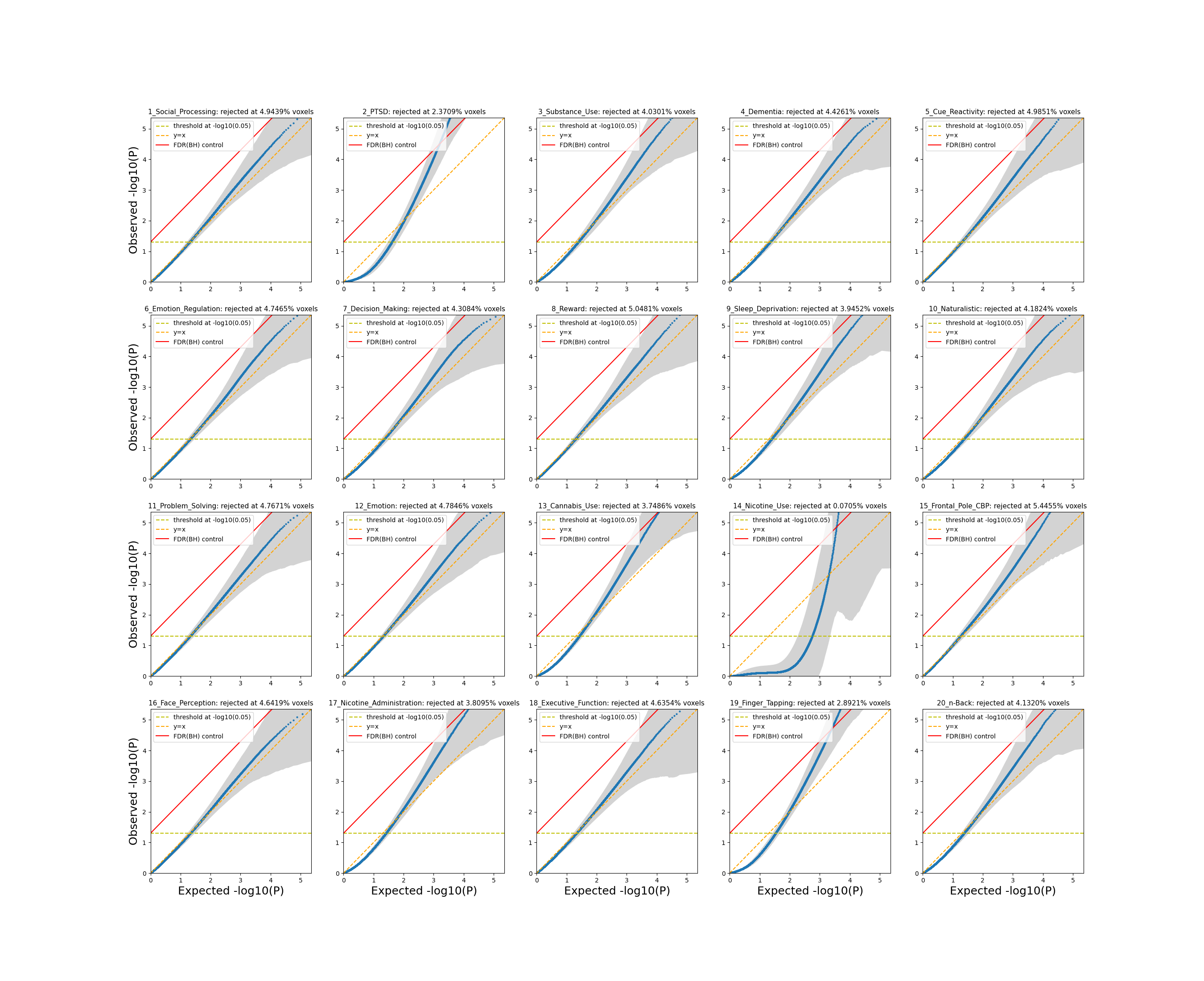}
\caption{P-P plot of $p$-value (under $-\log_{10}$ scale) with all of $20$ meta-analytic datasets, estimated by CBMR with Poisson model without study-level covariates, sampled with model-based approach.} 
\label{figure: full PP-plots}
\end{figure}

Previously, we have only displayed the PP-plots of spatial homogeneity tests of four representative datasets in \Cref{subsection: Simulation results}, here we will include all PP-plots on $20$ meta-analytic datasets in \Cref{figure: full PP-plots}.

\subsection{Likelihood-based comparison between Poisson, NB and clustered NB model}
\label{Appendix: Likelihood-based comparison}
\begin{figure}[h]
\centering
\includegraphics[width=16cm]{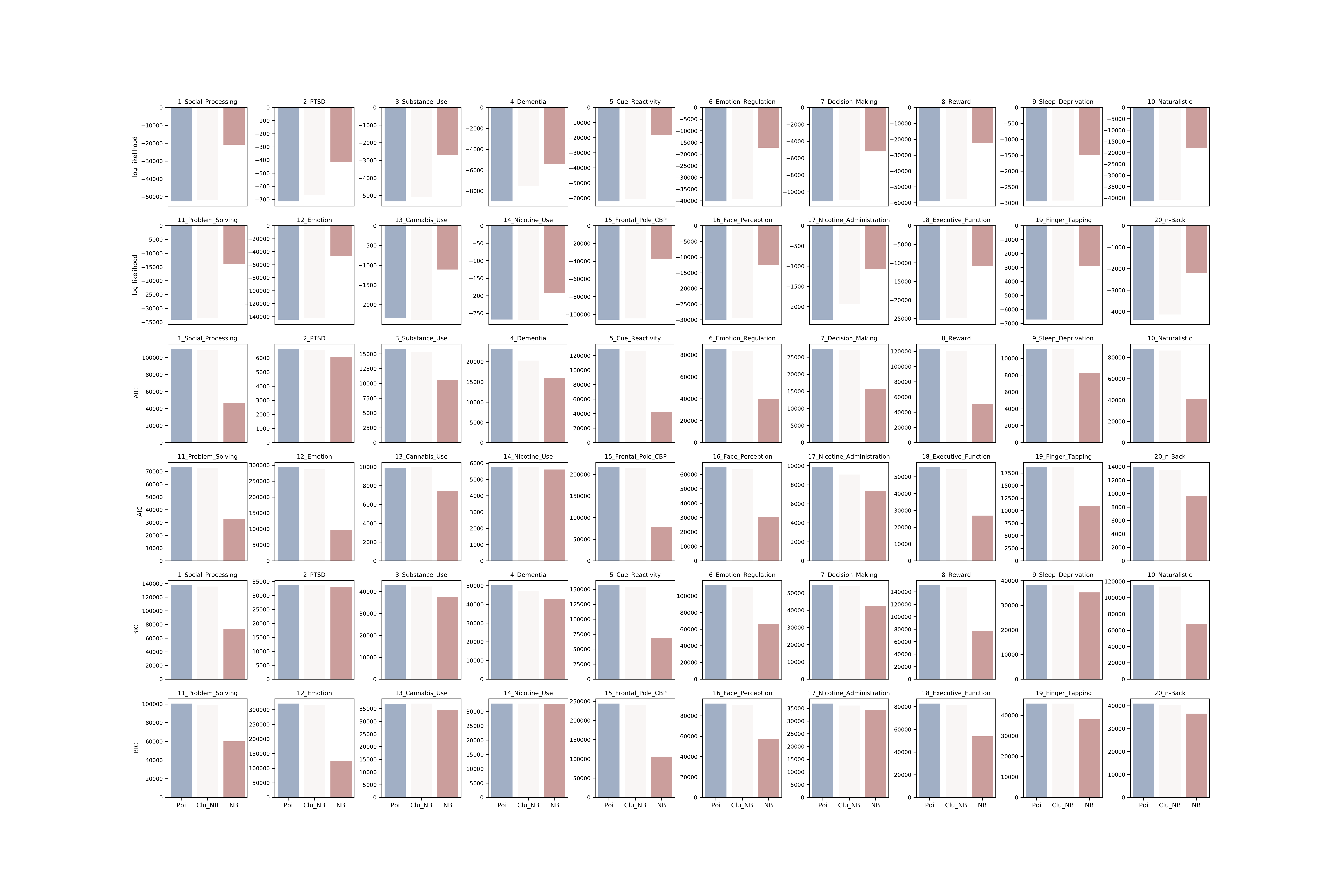}
\caption{Likelihood-based comparison (maximised log-likelihood, AIC and BIC) within CBMR with Poisson, NB and clustered NB model, across all of $20$ meta-analytic datasets. We found that NB model always gives the most accurate estimation, and existence of excess variance has been justified among CBMA data.} 
\label{figure: log_likelihood plot}
\end{figure}

To demonstrate the likelihood-based comparison between Poisson, NB and clustered NB model, we plot the maximised log-likelihood, AIC and BIC for each of the $20$ meta-analytic datasets in \Cref{figure: log_likelihood plot}. We also conduct Likelihood ratio test (LRT) to evaluate the trade-off between model sufficiency and complexity, here, we only list the p-values of LRT between Poisson and clustered NB model in \Cref{table: p-value for LRT}, as $p<10^{-8}$ for LRT between Poisson and NB model for each of the $20$ meta-analytic datasets.

\begin{table*}[!htb]
  \centering
  \footnotesize
  \begin{threeparttable}[]
  \caption{p-values of Likelihood Ratio test between Poisson and clustered NB model}
  \begin{tabular}{l|l|l|l|l|l}
    \toprule
    Dataset  & p-value & Dataset & p-value & Dataset & p-value \\
    \midrule
    Social Processing & $p<10^{-8}$ & PTSD & $p<10^{-8}$ & Substance Use & $p<10^{-8}$ \\
    Dementia & $p<10^{-8}$ & Cue Reactivity & $p<10^{-8}$ & Emotion Regulation  & $p<10^{-8}$ \\
    Decision Making & $p<10^{-8}$ & Reward & $p<10^{-8}$ & Sleep Deprivation & $p<10^{-8}$ \\
    Naturalistic & $p<10^{-8}$ & Problem Solving & $p<10^{-8}$ & Emotion & $p<10^{-8}$ \\
    Cannabis Use & $1$ & Nicotine Use & $p<10^{-8}$ & Frontal Pole CBP & $p<10^{-8}$ \\
    Face Perception & $p<10^{-8}$ & Nicotine Administration & $0.99$ & Executive Function & $p<10^{-8}$ \\
    Finger Tapping & $0.99$ & n-Back & $p<10^{-8}$ & \\
    \bottomrule
  \end{tabular}
  \label{table: p-value for LRT}
  \end{threeparttable}
\end{table*}

\subsection{Effect of study-level covariates}
Here, we investigate the effect of study-wise (square root) sample size and year of publication (after centring and standardisation) on each of $20$ meta-analytic datasets. Under the null hypothesis that regression coefficient of each covariates is not distinguishable from $0$ ($\gamma_i=0$ for $i=1,2$), we conduct hypothesis testing and summarise $Z-$score and $p-$value in \Cref{table: Hypothesis testing on effect of study-level covariates}.

\begin{table*}[t]  
  \centering
  \footnotesize
  \begin{threeparttable}[]
  \caption{Hypothesis testing on the effect of two study-level covariates on $20$ meta-analytic datasets}
  \tabcolsep=0.09cm 
  \begin{tabular}{l|cc|cc}
    \toprule 
     & \multicolumn{2}{c|}{(Square root) sample size} & \multicolumn{2}{c}{Year of publication}  \\
    \toprule
    Dataset & $Z-$score & p-value & $Z-$score & p-value \\
    \toprule
    Social Processing & $10.9053$ & $p<10^{-8}$ & $0.4164$ & $0.6771$ \\
    PTSD & $2.8789$ & $0.004$ & $0.6029$ & $0.5466$ \\
    Substance Use & $4.3887$ & $1.1404\times10^{-5}$ & $6.8398$ & $p<10^{-8}$ \\
    Dementia & $20.7177$ & $p<10^{-8}$ & $-1.3985$ & $0.1620$ \\
    Cue Reactivity & $6.1454$ & $p<10^{-8}$ & $-0.6880$ & $0.4915$ \\
    Emotion Regulation & $6.8934$ & $p < 10^{-8}$ & $-3.9588$ & $7.5329\times 10^{-5}$ \\
    Decision Making & $4.1104$ & $3.9499\times 10^{-5}$ & $0.1060$ & $0.9156$ \\
    Reward & $-0.1228$ & $0.9022$ & - & - \\
    Sleep Deprivation & $12.8765$ & $p<10^{-8}$ & $0.4201$ & $0.6744$ \\
    Naturalistic & $1.7038$ & $0.0884$ & $0.5395$ & $0.5896$ \\
    Problem Solving & $4.3079$ & $1.6485\times 10^{-5}$ & $2.2789$ & $0.0227$ \\
    Cannabis Use & $3.5915$ & $3.2878\times10^{-4}$ & $2.2117$ & $0.0270$ \\
    Nicotine Use & $5.0024$ & $5.6631\times10^{-7}$ & $3.1836$ & $0.0015$ \\
    Frontal Pole CBP & $5.5190$ & $3.4101\times10^{-8}$ & $7.4040$ & $p<10^{-8}$ \\
    Face Perception & $3.4090$ & $6.5212\times10^{-4}$ & $5.1018$ & $3.3651\times10^{-7}$ \\
    Nicotine Administration & $1.4594$ & $0.1445$ & $-1.0516$ & $0.2930$ \\
    Executive Function & $1.6989$ & $0.0932$ & $0.5047$ & $0.6138$ \\
    Finger Tapping & - & - & $0.1764$ & $0.8600$ \\
    n-Back & $1.4616$ & $0.1439$ & $0.1239$ & $0.9014$ \\
    \bottomrule
  \end{tabular}
  \vspace*{-0.1cm}
  \label{table: Hypothesis testing on effect of study-level covariates}
  \end{threeparttable}
\end{table*}
\end{document}